\documentclass[floatfix,aps,pra,superscriptaddress, longbibliography]{revtex4-2}
\usepackage{color}
\usepackage{xcolor}
\usepackage{amsmath}
\usepackage{amsthm}
\usepackage{amssymb}
\usepackage{graphicx,hyperref}

\newcommand{\CIE}{{\rm CIE}}
\newcommand{\CDE}{{\rm CDE}}
\newcommand{\tmut}{t_{\rm MUT}}

\begin{document}

\title{Emergent time scales of epistasis in protein evolution}

\author{Leonardo Di Bari}
\affiliation{DISAT, Politecnico di Torino, Corso Duca degli Abruzzi, 24, I-10129, Torino, Italy}

\author{Matteo Bisardi}
\affiliation{Sorbonne Universit\'{e}, CNRS, Institut de Biologie Paris-Seine, Laboratoire de Biologie Computationnelle et Quantitative -- LCQB Paris, France}

\author{Sabrina Cotogno}
\affiliation{Sorbonne Universit\'{e}, CNRS, Institut de Biologie Paris-Seine, Laboratoire de Biologie Computationnelle et Quantitative -- LCQB Paris, France}

\author{Martin Weigt}
\email{To whom correspondence should be addressed. E-mail: martin.weigt@sorbonne-universite.fr, francesco.zamponi@uniroma1.it}
\affiliation{Sorbonne Universit\'{e}, CNRS, Institut de Biologie Paris-Seine, Laboratoire de Biologie Computationnelle et Quantitative -- LCQB Paris, France}

\author{Francesco Zamponi}
\email{To whom correspondence should be addressed. E-mail: martin.weigt@sorbonne-universite.fr, francesco.zamponi@uniroma1.it}
\affiliation{Dipartimento di Fisica, Sapienza Universit\`a di Roma, Piazzale Aldo Moro 5, 00185 Rome, Italy}

\begin{abstract}
We introduce a data-driven epistatic model of protein evolution, capable of generating evolutionary trajectories spanning very different time scales reaching from individual mutations to diverged homologs. Our in silico evolution encompasses random nucleotide mutations, insertions and deletions, and models selection using a fitness landscape, which is inferred via a generative probabilistic model for protein families. We show that the proposed framework accurately reproduces the sequence statistics of both short-time (experimental) and long-time (natural) protein evolution, suggesting applicability also to relatively data-poor intermediate evolutionary time scales, which are currently inaccessible to evolution experiments. Our model uncovers a highly collective nature of epistasis, gradually changing the fitness effect of mutations in a diverging sequence context, rather than acting via strong interactions between individual mutations. This collective nature triggers the emergence of a long evolutionary time scale, separating fast mutational processes inside a given sequence context, from the slow evolution of the context itself. The model quantitatively reproduces epistatic phenomena such as contingency and entrenchment, as well as the loss of predictability in protein evolution observed in deep mutational scanning experiments of distant homologs. It thereby deepens our understanding of the interplay between mutation and selection in shaping protein diversity and novel functions, allows one to statistically forecast evolution, and challenges the prevailing independent-site models of protein evolution, which are unable to capture the fundamental importance of epistasis.
\end{abstract}

\maketitle

Proteins are essential molecules in living organisms. Their sequences have evolved over billions of years, giving rise to the impressive structural and functional variability we can observe today \cite{uniprot2023uniprot,burley2023rcsb}. Understanding this evolutionary process can shed light on the deep history of life on Earth and help researchers design novel proteins for various applications. One intriguing aspect of protein evolution is the possibility to conserve protein folds and function even when the protein sequence is mutated to reach a sequence divergence (i.e. percentage of mutated sites) as high as $\sim$70-80\%, thereby forming vast homologous families of sequence-diverged, but structurally and functionally highly similar proteins \cite{blum2021interpro}. On the other hand, a handful of random mutations in a protein may interrupt its functionality or destabilize its fold; the vast majority of mutations being deleterious \cite{fowler2014deep}. These observations illustrate the complex interplay between the stochasticity of mutations and functional selection, which are at the root of the diversity that Life has evolved.

A crucial concept in protein evolution is that of {\it epistasis}~\cite{de2014empirical,starr2016epistasis,johnson2023epistasis}: the effect of mutating an amino acid on a given position in a sequence depends on the {\it context}, i.e. on the amino acids that are present in the other positions of the sequence. 
The network of such epistatic interactions between sites has been investigated extensively, both experimentally~\cite{lunzer2010pervasive,biswas2019epistasis,poelwijk2019learning,domingo2019causes,park2022epistatic,chen2023understanding} and computationally~\cite{rivoire2016evolution,poelwijk2016context,reddy2021global,otwinowski2018biophysical,cocco2018inverse,vigue2022deciphering}. It is often found that pairwise interactions between mutations are weak, not even measurable in most cases.
Instead, epistasis emerges collectively by the accumulation of many weak interactions between a single site and many other sites along the protein sequence.

As an expected consequence of this observation, in most cases, the effect of any given mutation does not change dramatically when one or a few other mutations happen in the context. Only when a sufficient number of mutations is accumulated, 
one observes a substantial epistatic effect. While this intuition has been experimentally confirmed by comparing Deep Mutational Scans (DMS) 
of distant
homologs~\cite{lunzer2010pervasive,biswas2019epistasis,park2022epistatic,chen2023understanding}, its influence on the evolutionary dynamics has not been quantified because these experiments cannot track evolutionary trajectories.

Clarifying the impact of epistasis on evolutionary dynamics is also motivated by the recent explosion of experiments that investigate protein evolution {\it in vitro}, thanks to next-generation sequencing techniques \cite{joyce2007forty,matsuura2006vitro,fantini2020protein,stiffler2020protein,erdougan2023neutral}. A variety of experimental techniques has been developed for this purpose, such as directed evolution \cite{romero2009exploring}, in which mutations are artificially introduced in the target gene, fit mutants are selected, and the whole process is repeated until a fitness
optimum is reached~\cite{joyce2007forty,matsuura2006vitro}. Of particular interest for our work are neutral drift experiments, i.e. evolutionary experiments performed under weak selection, that can explore a large diversity of sequences in the fitness landscape around a given wild type~\cite{fantini2020protein,stiffler2020protein,erdougan2023neutral}. These laboratory techniques can potentially be compared with natural evolution, which also explores a wide diversity of sequences.

Yet, crucially, these time- and cost-intensive experiments are currently limited to rather short evolutionary trajectories. 
The diversity of the generated libraries is such that the average sequence divergence is at most about 20\%, hence much lower than that of natural libraries. A wide range of evolutionary time scales, located in between the short time scale of evolution experiments and the very long time scale of natural evolution, thus remains unexplored.

Characterizing quantitatively these rather data-poor intermediate time scales, and how epistatic effects emerge in this time regime, is the primary scope of the present work. Leveraging well-established data-driven fitness landscapes based on natural sequences belonging to entire protein families~\cite{morcos2011direct,figliuzzi2016coevolutionary,figliuzzi2018pairwise,russ2020evolution,barrat2021sparse,muntoni2021adabmdca}, we develop an in silico evolutionary dynamics in such a fitness landscape using a simple, biologically motivated stochastic process -- similar to ~\cite{de2020epistatic,bisardi2022modeling,alvarez2024vivo,doi:10.1073/pnas.2316662121,biswas2024kinetic}, but taking explicitly into account mutational constraints due 
to the genetic code, and the possibility of insertions and deletions.
Consistently with~\cite{bisardi2022modeling}, 
using as the only input divergent natural (around $70 \%$ of amino acids are different) sequences within the same family,
the in silico evolution reproduces very well the short-time evolution sequencing data from two distinct experiments. 
On such short times, we confirm epistatic effects to be scarce~\cite{park2022epistatic}: a non-linear function of an additive fitness based on a single-mutant DMS is a good predictor of the effect of multiple mutations~\cite{otwinowski2018biophysical,bisardi2022modeling}. 
At very long time scales, by construction, our in silico evolution samples sequences in equilibrium, i.e., model-generated sequences at long times have the same statistical properties of natural sequences, and epistasis is extremely relevant~\cite{russ2020evolution}. In other words, our model is {\it generative}, and reproduces the statistics of both short-time experimental and long-time natural evolution within a single framework.

Our most important results are the following. (i)~We show that epistasis clearly emerges on intermediate scales, where sequence divergence is ${\approx~40-50}$\%, i.e. much higher than that of evolution experiments, but much lower than that of natural sequence homologs.
Over these time scales, a sufficient number of mutations accumulate, such that epistatic effects are prevalent and deeply influence the evolutionary dynamics.
(ii)~We introduce a classification of sites as `variable', `conserved', and `epistatic', solely based on the model-predicted fitness around a specific reference protein (e.g.~the wild type in experimental evolution). We show that this classification is predictive of future evolution. In fact,
variable sites display fast evolutionary dynamics almost without epistasis. Conserved sites only rarely mutate, and yet they are not constrained by epistasis but rather by functional properties.
Epistatic sites show a high context-dependence, and their dynamics also shows non-trivial, highly collective features on intermediate time scales, making their evolution hardly predictable.
(iii)~We show that contingency and entrenchment effects appear on the same intermediate time scales. 
As shown in the experiment of Ref.~\cite{park2022epistatic},
they are manifested by the decrease of the fitness of a given mutation both before and after the time at which it is introduced. We show that
the model can reproduce the experimental trends of fitness variation.

In summary, our data-driven modeling approach is able to quantitatively simulate protein evolution over the full range of time scales, thus enhancing our understanding of this complex multi-scale process itself,
and as such it can provide important guidance for new experiments. For example, one might use the model to design new in vitro evolution 
protocols in which the selection is varied over time in such a way to optimize the diversity and functionality of the final library~\cite{erdougan2023neutral}, or to study mutational paths
between distant homologs~\cite{poelwijk2019learning,tian2020exploring,phillips2021binding,mauri2023mutational} and the emergence of new functions.

\begin{figure*}[t]
\centering
\includegraphics[width=.85\linewidth]{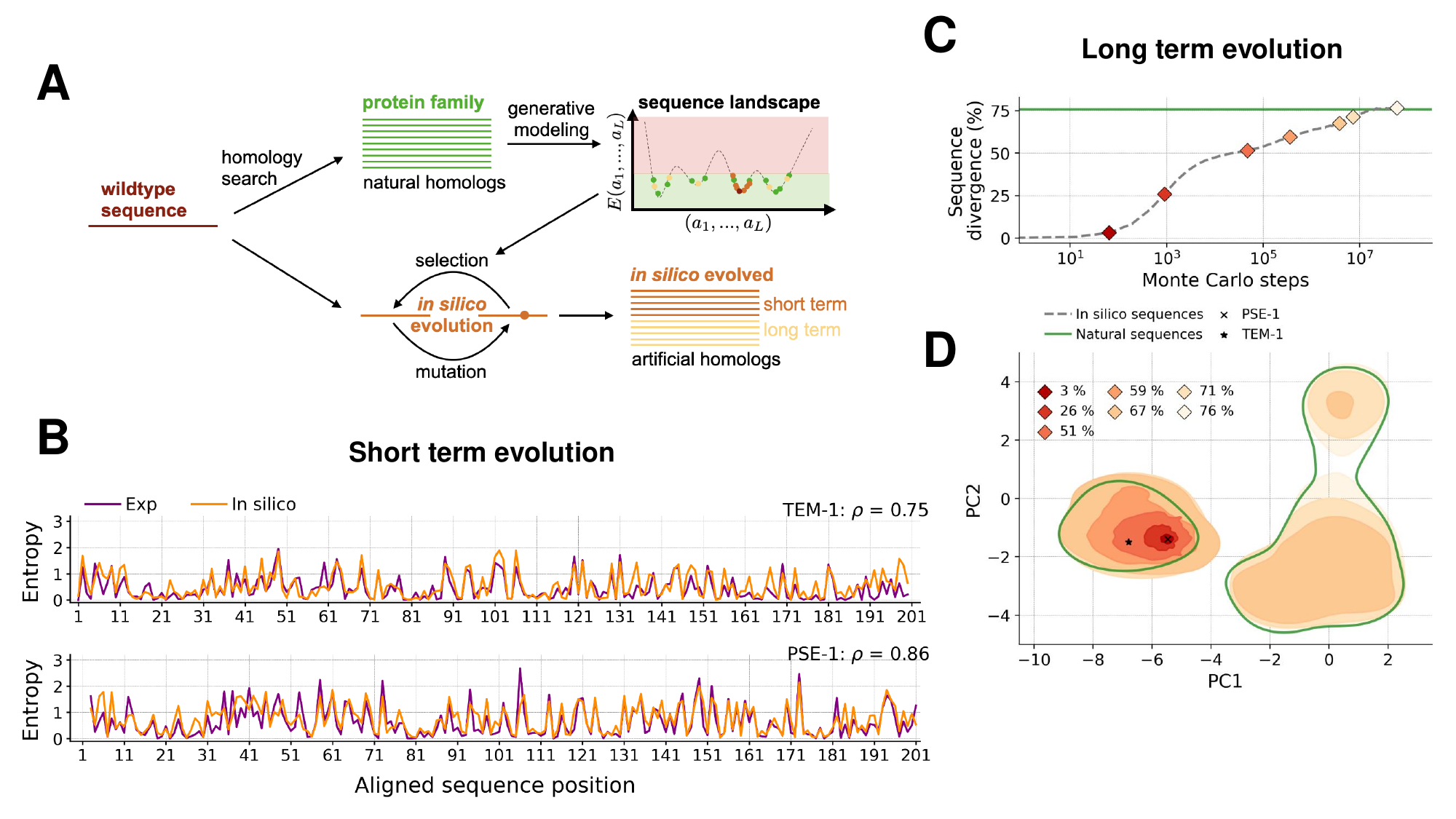}
\caption{
{\bf Illustration and validation of the evolutionary model.}
(A) Schematic description of the MCMC simulation procedure adopted to mimic the evolutionary dynamics.  (B) Reproducing the short-term in vitro evolution of the PSE-1 and TEM-1 $\beta$-lactamase 
wild types: comparison of experimental (violet) and in silico (orange) site entropies. (C,D) Reproducing the long-term evolution of the $\beta$-lactamase family. (C)
Average sequence divergence from the wildtype(normalized Hamming distance) over 1000 Markov chain realizations of the mutational dynamics at $T=1$,
compared to the average divergence from the wildtype of the family of natural proteins used as a training set. (D) Projection over the two principal components (PCA) of the natural proteins (green) and
of the sequences sampled after a given number of in silico evolutionary steps (red to orange, labeled by the average sequence divergence).}
\label{fig:1}
\end{figure*}

\section*{A quantitative epistatic model bridges timescales from short to long-term evolution}

Our general approach to modeling the evolution starting in some wild type (WT) protein is illustrated in Fig.~\ref{fig:1}A.

First, we collect a multiple-sequence alignment (MSA) of possibly very diverged, but naturally occurring homologs, i.e.~of the protein family sharing evolutionary ancestry with the WT, cf.~{\em Methods}. The MSA serves to estimate a data-driven fitness landscape $F(\mathbf{a})$ of aligned amino acid sequences $\mathbf{a}$ via 
Direct Coupling Analysis (DCA) using
Boltzmann Machine (BM) learning~\cite{figliuzzi2018pairwise}, which aims at assigning high probabilities $P(\mathbf{a})$ to functional sequences, and low probabilities to non-functional ones.
In this work we used the bmDCA 
 implementation of Ref.~\cite{muntoni2021adabmdca}. 
 Upon sampling sequences from the inferred $P(\mathbf{a})$, the model correctly reproduces many statistical features of the input MSA, such as the distribution of sequence divergences, amino acid frequencies (i.e. conservation) and pairwise covariations (i.e. coevolution)~\cite{figliuzzi2018pairwise}. 
The model is informative at very different divergence scales: at the scale of individual mutations, it allows to predict the outcome of DMS experiments~\cite{figliuzzi2016coevolutionary,barrat2021sparse}, even for distant homologs~\cite{chen2023understanding}, to forecast viral evolution~\cite{rodriguez2022epistatic,biswas2024kinetic}
and to detect selection acting on polymorphisms~\cite{couce2017mutator,vigue2022deciphering,vigue2023predicting}. On a global scale, it is a generative model: subsets of artificial sequences sampled from $P(\mathbf{a})$ display comparable functionality with natural ones~\cite{tian2018co,russ2020evolution, alvarez2024vivo}. This proven capacity to bridge scales in sequence divergence is a major motivation for us to use the bmDCA model as an evolutionary fitness landscape. 

Technically, the model assigns a statistical energy $E(\mathbf{a})$ to each sequence $\mathbf{a}=(a_1,\cdots,a_L)$, where the $a_i$ describe the $20$ amino acids plus a gap symbol. The DCA energy is mathematically given (up to a constant) by ${E(\mathbf{a})=-\log P(\mathbf{a})}$. It is also biologically interpreted as being related to a given, measurable, fitness by a non-linear relationship
${F(\mathbf{a}) = \phi(E(\mathbf{a}))}$~\cite{morcos2014coevolutionary,barrat2016improving,chen2023understanding}, with $\phi(E)$ a non-linear function that we discuss below.

Second, we model evolution (natural and experimental) as a stochastic process in this energy landscape, with random mutations and phenotypic selection modeled by $E(\mathbf{a})$~\cite{de2020epistatic}; simulating an evolutionary trajectory becomes a Markov Chain Monte Carlo (MCMC) process initiated in the WT sequence. Because the number of MCMC steps has no direct biological meaning, the MCMC process is iterated until the desired number of mutations is reached. Taking a step forward with respect to previous work~\cite{de2020epistatic,bisardi2022modeling,alvarez2024vivo,doi:10.1073/pnas.2316662121},
our biologically motivated mutational process uniquely integrates several crucial requirements into a unified framework, providing a consistent and quantitative description across time scales:
\begin{enumerate}
    \item single nucleotide mutations, insertions and deletions, which are then translated into amino acid changes through the genetic code;
    \item selection via the data-driven and epistatic landscape learned from natural sequences, with the possibility to modulate selection strength;
    \item detailed balance to ensure convergence to the natural distribution of the training dataset at long times.
\end{enumerate}
While a mutational model defined directly on amino acids captures many consequences of epistasis on protein evolution~\cite{de2020epistatic,doi:10.1073/pnas.2316662121}, a more quantitative model requires to respect the biological nature of mutations, insertions, and deletions acting at the nucleotide level~\cite{bisardi2022modeling,alvarez2024vivo,gunnarsson2023predicting}, especially when trying to model experimental results at short time scales. However, while the evolutionary dynamics of~\cite{bisardi2022modeling} is restricted to model short-time evolution, here we design a novel mutational process that mathematically guarantees convergence to the correct asymptotic sequence distribution, represented by the natural MSA, cf.~{\em Methods} and the SI.
These dynamics are, in principle, applicable to arbitrary evolutionary timescales, from the emergence of the first mutation to divergent evolution over billions of years.

The DCA model is trained on amino acid sequences, hence when converting it to the nucleotide space, the probability is uniformly distributed among all possible nucleotide sequences coding for the same amino acid sequence, without taking into account the codon bias. We checked that including codon bias does not change the results of our analysis. Deletions and insertions are also necessary to recover the natural data statistics at long time scales (SI, Fig.~S10). Furthermore, the selective pressure is modeled through a `selection temperature', i.e.~a global rescaling of the statistical energy $E(\mathbf{a}) \to E(\mathbf{a})/T$, which allows us to reproduce different experimental conditions~\cite{bisardi2022modeling}:  low (high) temperature corresponds to high (low) selection, because the higher the temperature, the bigger the variance of the mutational effects accepted by the model. 

We validate our new MCMC process by comparing our simulations with both experimental evolution (Fig.~\ref{fig:1}B) and natural datasets (Fig.~\ref{fig:1}C-D) of protein sequences belonging to the $\beta$-lactamase family, representing respectively short- and long-term evolution. 
In vitro evolution was performed in~\cite{stiffler2020protein,fantini2020protein} at moderate selection strength, by growing mutant strains in a medium containing a concentration of antibiotic that is modulated around the wild type EC$_{50}$. This allows for the generation of a diverse library of mutant strains having comparable resistance to the wild type~\cite{stiffler2020protein,fantini2020protein,erdougan2023neutral}.
These datasets are compared to in-silico data, generated by simulating many MCMC trajectories starting from a reference WT and performing a fixed number of MCMC steps $n$. 

In short-term evolution, we fix $n$ and the selection strength $T$ such that the average number of mutations and the average model energy $E(\mathbf{a})$ of the simulated MSA match those of the experimental libraries of~\cite{stiffler2020protein,fantini2020protein}, cf.~{\em Methods}.
Next, in each of the in-silico and experimental libraries, we compute the frequencies $f_i(a)$ of observing amino acid $a$ at site $i$, and determine the resulting position-specific {\it entropies} 
\begin{equation}\label{eq:Si}
S_i = -\sum_a f_i(a) \log_2 f_i(a)\ ,\quad i=1,\dots,L\ .
\end{equation}
A fully conserved site implies $S_i=0$, while a fully mutable site (all amino acids and gap equiprobable) would lead to $S_i = \log_2 21 \sim 4.39$.
In general, ${\cal N}_i = 2^{S_i}$ represents an `effective number' of viable amino acids at site $i$.
Fig.~\ref{fig:1}B shows a remarkable agreement between the entropies predicted by the model and those observed experimentally, for evolutionary trajectories starting from two distinct $\beta$-lactamase enzymes, TEM-1 in~\cite{fantini2020protein} and PSE-1 in~\cite{stiffler2020protein}.
It is important to underline that the high correlation between model and experiments is far from obvious, considering that the landscape $E(\mathbf{a})$ was learned on natural sequences that are diverged from both TEM-1 and PSE-1, hence the training MSA has no explicit information about the local fitness landscape around these WTs.
Also, the model is able to reproduce the local entropy on $\sim 200$ distinct sites around two distinct wild types, while having only two free parameters $n$ and $T$.

Figs.~\ref{fig:1}C-D analyze the long-time evolution of our model. When increasing the MCMC steps, the statistics of the model-generated sequences slowly converges to that of the natural MSA of the $\beta$-lactamase family (represented in green in both figures).
In particular, Fig.~\ref{fig:1}C shows the average distance from the wildtype for sequences in the simulated MSA, and Fig.~\ref{fig:1}D shows the spreading of sequences in the two-dimensional space defined by the two principal components of the natural sequences. The asymptotic agreement is expected by construction, because (i) our MCMC dynamics satisfies detailed balance with respect to the model probability $P(\mathbf{a})\propto \exp[- E(\mathbf{a})/T]$, which guarantees that the long-time dynamics samples sequences independently from $P(\mathbf{a})$, and (ii) previous studies have shown that $P(\mathbf{a})$ obtained from bmDCA is a good generative model~\cite{figliuzzi2018pairwise,russ2020evolution,barrat2021sparse, alvarez2024vivo}. For further proof of the generative quality of the model, it is shown in SI, Fig.~S1, that generated sequences correctly reproduce the statistics of the training data.
Yet, the way in which the MCMC dynamics converges to equilibrium is far from trivial: after an initially fast accumulation of mutations ($\gtrsim 25\%$ in $10^3$ MCMC steps), we observe the emergence of a {\it plateau} of sequence divergence at intermediate times, such that the initial evolutionary dynamics remains confined in a limited neighborhood of divergence $\lesssim 50\%$ from the starting sequence. 
Only at much larger times, our dynamics explores the full space of natural sequences, reaching divergences above $50\%$ and close to the value of $\sim 75\%$ found in the natural MSA.
The same phenomenology is observed in Fig.~\ref{fig:1}D: on intermediate time scales the dynamics remains confined in the PCA cluster in which it was initialized, and only at much longer times it is able to jump to the other clusters. We note that this behavior of the sequence divergence is specific to the $\beta$-lactamase family, and other families may display a different evolutionary dynamics depending on their fine organization into subfamilies~\cite{barrat2021sparse}, cf. SI, Fig.~S2. 
However, we note that this complex evolutionary dynamics is possible only in an epistatic model; a model with a non-epistatic, additive fitness landscape shows a simple dynamics characterized by a single time scale of sequence divergence (SI, Fig.~S6).

Our data-driven model is thus able to connect time scales spanning from a few mutations ($\sim$ 1 year, cf.~the evolution of SARS-CoV2 variants)~\cite{rodriguez2022epistatic} up to distant homology ($\sim 10^9$ years, i.e. the timescale of life on Earth) in a quantitative way. 
We note that an evolutionary model capturing such fine-scale features in completely different contexts is quite rare: 
for instance, most phylogeny and ancestral sequence reconstruction algorithms are based on independent-site evolution and therefore neglect epistasis~\cite{felsenstein}. 
Our model thus has the potential to contribute to a deeper understanding of the interplay between mutations and selection in shaping protein diversity, leading to the emergence of new protein functions.

\begin{figure*}[t]
\centering
\includegraphics[width=.8\linewidth]{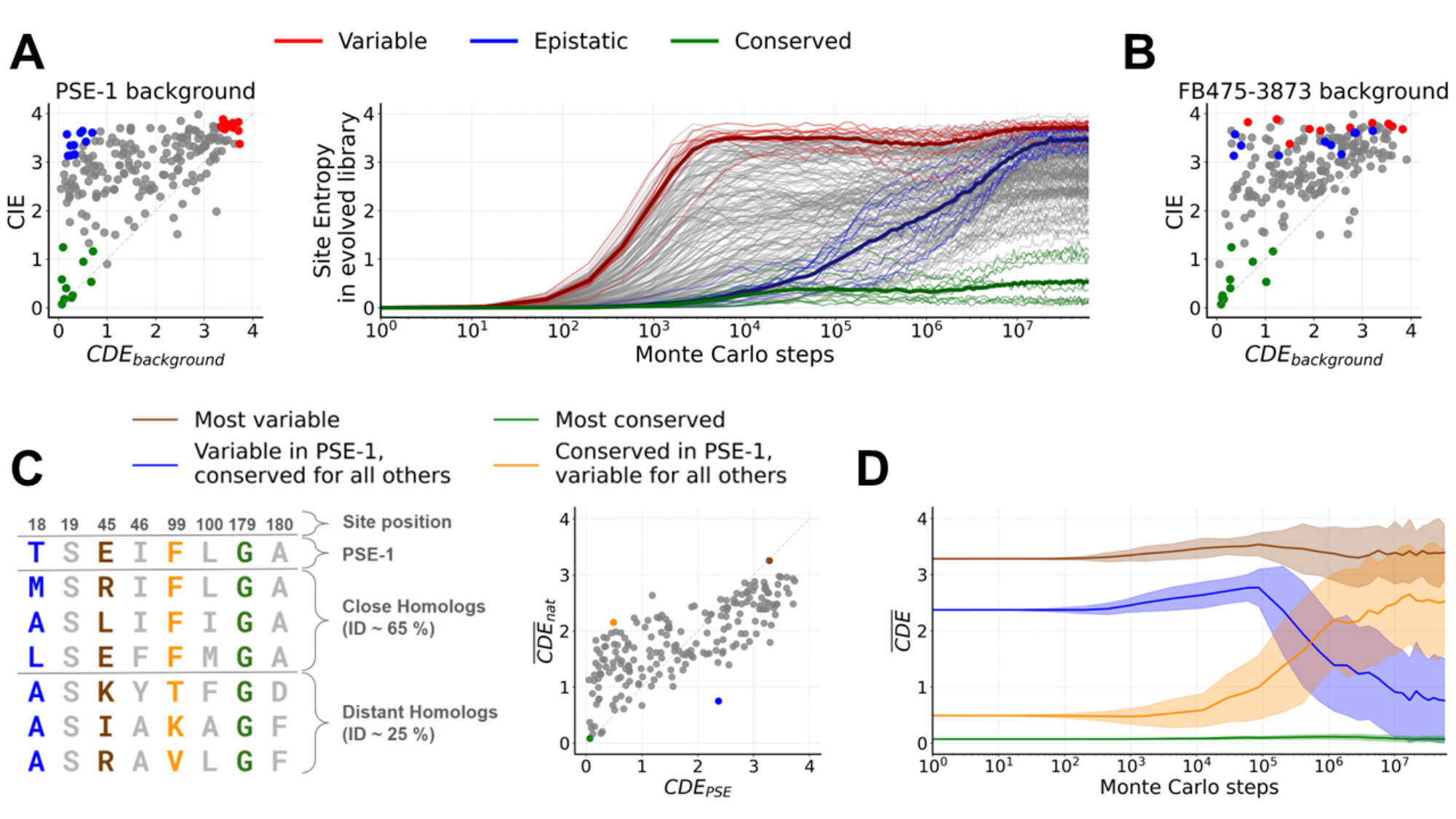}
\caption{
{\bf In silico evolutionary dynamics of variable, conserved, and epistatic sites for the $\beta$-lactamase family.}
(A) Left: classification of residues according to local variability (CDE) computed with the bmDCA model in the PSE-1 context and global variability (CIE) computed using the amino acid frequencies in natural proteins. We identify three categories of 10 sites each: conserved (green), mutable (red) and epistatic (blue). Right: Site entropy of a library constructed by 1000 independent MCMC samplings initialized in PSE-1, with the three different site categories highlighted; faded lines refer to each site whereas the three thick lines are the mean over the three categories. (B) Same scatter plot as in left panel of A, but in the context of a distant homolog. The site colors are the same as for PSE-1, making clear that sites categories are reshuffled in a different context. (C) Illustration of four prototypical sites in the alignment and scatter plot of their CDE computed in the PSE-1 context versus the mean CDE on all proteins of the $\beta$-lactamase family. (D) CDE of the four sites identified in panel C, averaged over the context generated by 100 independent MCMC trajectory (thick lines). The faded area indicates the standard deviation.}
\label{fig:2}
\end{figure*}

\section*{Epistasis drives the emergence of long evolutionary timescales}

Previous studies suggest that epistasis (or context dependence) is a collective phenomenon that builds up when mutations accumulate~\cite{lunzer2010pervasive,biswas2019epistasis,park2022epistatic,chen2023understanding,vigue2022deciphering}. A statistical characterization of the evolutionary paths relating homologous proteins at different evolutionary distances would then be particularly useful for unveiling the origin and dynamics of epistatic effects during evolution. 
After having validated our model against data from short and long-term evolution, we can now use it to generate such trajectories and explore sequence space beyond experimental limits.
In particular, we can explore all intermediate time (and sequence divergence) scales between in vitro experiments and natural evolution, for which experimental data is currently lacking.

 Our model suggests that the mutational dynamics of sites is linked to their degree of epistasis, which we now quantify by taking inspiration from previous works on the E. coli genome~\cite{vigue2022deciphering} and the SARS-CoV-2 spike protein~\cite{rodriguez2022epistatic}.
We classify each site according to two distinct notions of entropy as a simple information-theoretic measure of mutability. The context-independent entropy (CIE) is estimated from the empirical frequencies $f^{\rm nat}_i(a)$ of amino acids $a$ in column $i$ of the MSA of natural sequences :
\begin{equation}\label{eq:CIE}
\CIE_i = -\sum_a f^{\rm nat}_i(a) \log_2 f^{\rm nat}_i(a) \ .
\end{equation}
The quantity reflects the global mutability (as a complement of conservation) of site $i$ as observed across all natural sequences. Because bmDCA is generative, $f^{\rm nat}_i(a)$ is very close to the model marginal probability $p_i(a)$ of observing amino acid $a$ in position $i$.
Hence, the CIE expresses the site mutability in equilibrium, independently of the context.
The context-dependent entropy (CDE), instead, expresses the site mutability in a given amino acid context~\cite{vigue2022deciphering}.
 We consider the conditional probability $P(a_i=a | \mathbf{a}_{\setminus i})$ assigned by the bmDCA model to all the mutations on site $i$ within a given context $\mathbf{a}_{\setminus i} = (a_1,\cdots,a_{i-1},a_{i+1},\cdots,a_L)$ (i.e. sequence $\mathbf{a}$ without site~$i$). 
 From these conditional probabilities, we obtain
\begin{equation}
\CDE_i(\mathbf{a}_{\setminus i})
= -\sum_a P(a_i=a | \mathbf{a}_{\setminus i})
\log_2 P(a_i=a | \mathbf{a}_{\setminus i}) \ .
\end{equation}
Note that, differently from the CIE, the CDE depends explicitly on the chosen context. It is not given directly by the natural sequences, but requires the inference of a statistical model. It can also be estimated from an experimental DMS around the given reference context - a route we will not follow in this work, see Ref.~\cite{chen2023understanding}.

Fig.~\ref{fig:2}A shows a scatter plot of the CIE versus the CDE for each site in the $\beta$-lactamase PSE-1 wild type context. The plot
identifies three relevant subcategories of sites.
Variable sites (high CIE and CDE, red in Fig.~\ref{fig:2}A) are very variable in the natural MSA and mutable in the PSE-1 context, hence they are not strongly constrained
by the context. Conserved sites (low CIE and CDE, green in Fig.~\ref{fig:2}A) hardly mutate, as they are conserved both in the MSA and in the PSE-1 context. 
Epistatic sites (high CIE and low CDE, blue in Fig.~\ref{fig:2}A) are strongly conserved within the PSE-1 context, but when the context is allowed to vary, they show high mutability, leading to high variability across homologs of the MSA.

To understand how these categories influence the evolution of individual positions in a protein sequence, and how these positions accumulate variability, we simulate a library of 1000 independent trajectories, all identically initialized in PSE-1. Taking the library at $n$ MCMC steps, we compute the site entropies $S_i(n)$ using~\eqref{eq:Si}. In Fig.~\ref{fig:2}A we plot, for each site, these entropies $S_i(n)$ as a function of $n$; the coloring evidencing the different categories.
For variable sites, we observe that after a short time ($n\sim 10^3$ steps, sequence divergence $\sim 25\%$) the site entropies reach their asymptotic values. For variable sites, this corresponds to $S_i(n) \sim \CIE_i \sim \CDE_i$. In other words, variable sites can mutate essentially independently of the context, in such a way that the 1000 evolutionary trajectories quickly display all possible mutations admitted on those sites.
For conserved sites, as expected, we only rarely see mutations, such that the site entropy remains very low at all times.
The most interesting behavior is that of epistatic sites. On short time scales, within their essentially unchanged context, these sites access only a small subset of mutations, which are compatible with the current context. Hence, on the short time scale $n\sim 10^3$ we have \[ S_i(n) \sim \CDE_i(\text{PSE-1})<\CIE_i \ : \] epistatic sites behave similarly to the conserved sites. However, on longer time scales ($n\sim 10^5 - 10^6$, sequence divergence $\sim 50\%$) enough mutations start to accumulate in the context, such that these sites collectively unlock and their entropy increases, asymptotically reaching their equilibrium value ${S_i(n\to\infty)\sim \CIE_i}$. This epistatic process happens in a wide range of time scales, spanning almost two orders of magnitude,
for different epistatic sites. It is possible that the unlocking of a mutation in one epistatic site allows other epistatic sites to mutate, hence leading to avalanches of mutations; we leave a finer characterization of these processes to future work.

\begin{figure*}[t]
\centering
\includegraphics[width=0.7\linewidth]{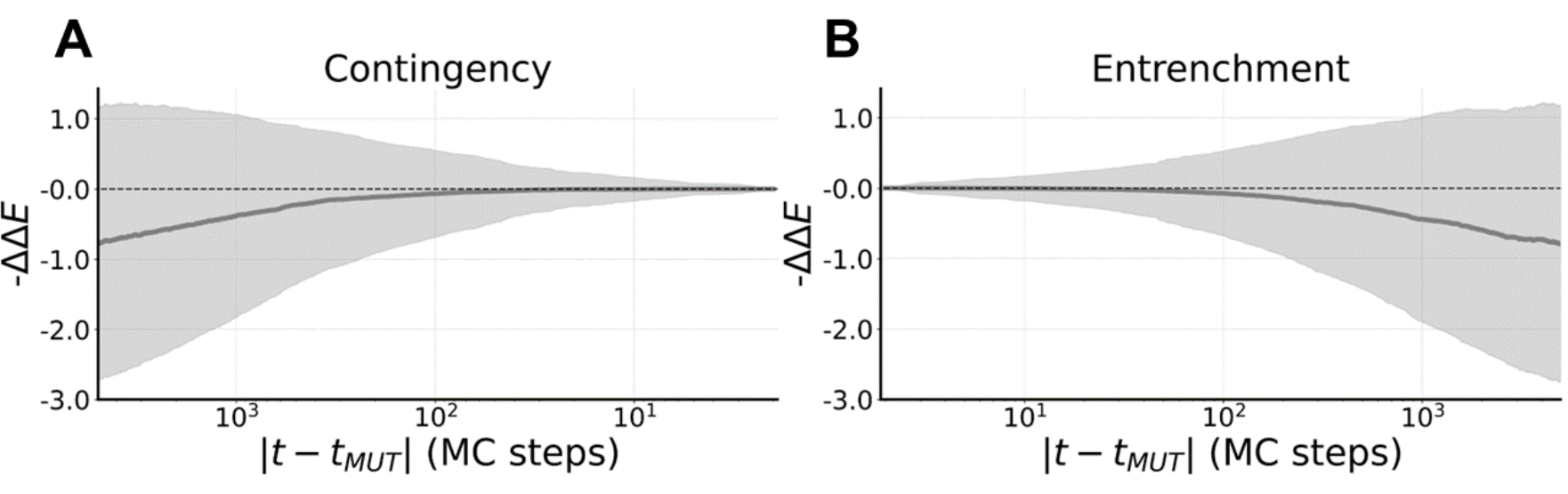}
\caption{
{\bf Contingency and entrenchment from in silico evolution for the $\beta$-lactamase family.}
We select 5000 mutations labeled by $aib$ (i.e. site $i$, amino acid $a\to b$) along an evolutionary trajectory of the $\beta$-lactamase family.
(A)~Contingency: average value of the fitness gain $-\Delta \Delta E$ between the original mutation $aib$ at time $\tmut$, and the mutational effect of introducing amino acid $b$ in site $i$ over a time window of 5000 MCMC steps before $\tmut$.
The faded area is the standard deviation. (B)~Entrenchment: 
average value of the fitness gain $-\Delta \Delta E$ between the original mutation $aib$ at time $\tmut$,
and the mutational effect of reverting to amino acid $a$ over a time window of 5000 MCMC steps after $\tmut$. The faded area is the standard deviation.
}
\label{fig:3}
\end{figure*}

It is important to stress once again that the CDE, hence the classification discussed above, depends on the context. Fig.~\ref{fig:2}B shows that in a distant context, randomly selected from the natural MSA, the CDE of the same sites shown in Fig.~\ref{fig:2}A has changed (note instead that the CIE does not change by construction), in such a way that variable and epistatic sites are reshuffled (note that instead conserved sites always remain conserved). This is also illustrated in Fig.~\ref{fig:2}C, where a scatter plot over sites of the average CDE over all natural contexts is shown against the CDE in the PSE-1 context. For illustration, a small MSA is also shown, containing a few selected sites and a few sequences at different sequence divergence from PSE-1.  
Fig.~\ref{fig:2}D shows, for the few selected prototypical sites, the dynamical evolution of the CDE averaged over the library of contexts obtained from 100 independent MCMC, all originating from PSE-1.
Consider as a first example the most conserved site in the natural library (green in Fig.~\ref{fig:2}C-D); because this site is conserved in all contexts, its CDE always remains low, even when the context is allowed to mutate. On the contrary, for the most variable site (brown in Fig.~\ref{fig:2}C-D), the CDE always remains high because it is high in all contexts.
More interestingly, we consider a site that has lower CDE in PSE-1 than on average (yellow in Fig.~\ref{fig:2}C-D). For this site, we observe that the CDE grows over time on the time scale $n\sim 10^5 - 10^6$, while the context mutates, reaching its average value asymptotically. Similarly a site with higher CDE in PSE-1 than on average (blue in Fig.~\ref{fig:2}C-D) displays a decreasing CDE over the same time scale. To further validate our observations, we repeat the same procedure for the DNA Binding Domain family (DBD), obtaining similar results (SI, Figs.~S2, S3, S4).

Overall, these findings show that epistasis manifests itself as a collective phenomenon~\cite{park2022epistatic} on intermediate time scales, much longer than those accessible by present-day in vitro evolutionary experiments, but much shorter than natural evolution. Note that, without the action of epistasis and thus without the possibility of emergent collective phenomena, all sites should evolve on identical short timescales, comparable to the one of the variable sites in our model (SI, Fig.~S6).

\section*{Context dependence shapes evolutionary paths: contingency and entrenchment}

Epistasis implies that the effect of a mutation is influenced by its context, i.e.~the rest of the sequence. It shapes evolutionary trajectories in two ways, denoted as contingency and entrenchment~\cite{starr2018pervasive}. Contingency means that a mutation happening (and surviving selection) at a given time, is contingent on previous mutations on other sites that make it possible (i.e.~not counter-selected against). Entrenchment means that a mutation, which typically has to be neutral at the time of its introduction to survive selection, can subsequently be entrenched by later mutations in other positions, such that its reversion becomes deleterious. 
Our work, consistently with recent experiments~\cite{park2022epistatic,chen2023understanding, 03e7c71a26cd4e26b3717cbb16d1505b}, suggests that both phenomena are caused by the accumulation of many mutations in the sequence context, each having a small epistatic impact, rather than just a few mutations with very strong epistatic effects. 

To further test this idea we first consider the negative variation of energy, $-\Delta E(t=\tmut)$, associated to a given mutation at the time $\tmut$ at which it is first observed along an evolutionary trajectory. Fitness is negatively correlated to energy, such that a positive $-\Delta E$ indicates a beneficial mutation.
We find that along the evolutionary trajectories, the distribution of $-\Delta E(t=\tmut)$ is centered close to zero. This means that observed mutations are close to neutral (SI, Fig.~S8), as opposed to a strongly deleterious bias of random mutations (remember that our evolutionary model combines mutation and selection in its elementary MCMC step). This is consistent with the DCA energy remaining on average constant along trajectories, 
but also with the empirical observation that natural amino acid polymorphisms tend to be neutral~\cite{vigue2022deciphering} when scored by a DCA model. 

To study contingency and entrenchment, following Ref.~\cite{park2022epistatic}, we consider the variation of the effect of a mutation $a\to b$ observed in site $i$ at time  $\tmut$, when either introduced at some time $t+\tmut$ prior to $\tmut$, or reverted at some later time $t+\tmut >\tmut$:
\begin{equation}\begin{split}
-\Delta\Delta E(t) &= -\Delta E(t+\tmut) + \Delta E(\tmut) \ .
\end{split}\end{equation}
The mutational effects are calculated for site $i$, but considering the different backgrounds present at $\tmut$ and $t+\tmut$. Once again, the sign is chosen in such a way that a positive $-\Delta\Delta E$ indicates a gain of fitness. 
For negative $t$, i.e.~times prior to the actual occurrence of the mutation $a\to b$, we test the insertion of amino acid $b$ into the past sequence $A(t+\tmut)$. Fig.~\ref{fig:3}A shows that the average $-\Delta\Delta E$ is negative when $t<0$ indicating a fitness loss, i.e. a mutation is typically deleterious at previous times, leading to contingency. For positive $t$, i.e.~times posterior to the actual occurrence of the mutation $a\to b$, we revert the amino acid present in $A(t+\tmut)$ to $a$. Fig.~\ref{fig:3}B shows that this reversion is typically deleterious at subsequent times, leading to entrenchment. 

Remarkably, both contingency and entrenchment start to manifest at time scales $|t|\sim 10^3$ MC steps, i.e.~at times where sites classified as epistatic in Fig.~\ref{fig:2}A are still conserved. In this sense, contingency and entrenchment can be interpreted as the first indicators of the changing context, which only at later times adds up to substantial changes in the context, and free epistatically constrained site for mutation. The same trend can also be seen in the DBD family (SI, Fig.~S5).

\begin{figure*}[t]
\centering
\includegraphics[width=.85\linewidth]{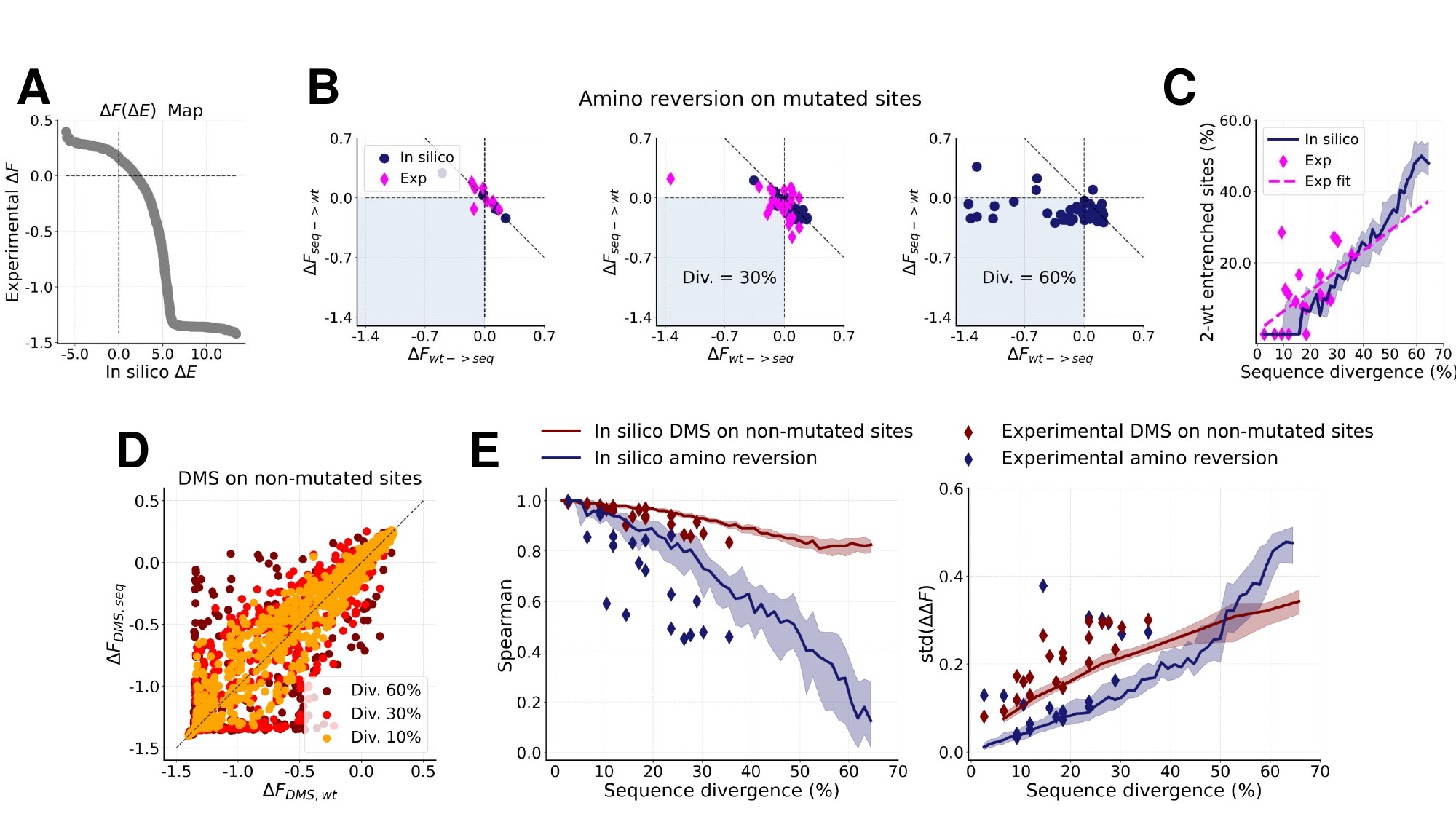}
\caption{
{\bf Comparison of in silico and experimental data for single mutations in distant homologs in the DBD family.}
(A) Visualization of the mapping $\Delta F = \phi(\Delta E)$ that is used to convert DCA energy into experimental
fitness.
(B) Scatter plots of the effect of mutating an amino acid of a wild type into the one of an evolved sequence versus the reversed mutation, both for experimental (magenta) and in silico (blue) sequences. Different panels show results for sequences at increasing sequence divergence from the wild type. 
Points that fall in the shaded area display sign epistasis, i.e. both the forward and reversed mutations are deleterious in their respective contexts.
(C) Percentage of mutated sites displaying sign epistasis (i.e. falling in the shaded areas in panels B) as a function of the evolutionary distance from the wild type for experiments (magenta) and simulated trajectories (blue). Dashed magenta line is a fit of the experimental points.
(D) Scatter plots of the mutational effects of all single amino acid variants (i.e. in silico DMS) around the evolved sequences versus the wild type, for sites that are not mutated between the two sequences. Different colors correspond to sequences at increasing divergence from the wild type. (E) Left: (minus) Spearman correlation of the amino acid reversions shown in panel B
for increasing sequence divergence, 
both for experimental and in silico sequences (blue curve and points).
Spearman correlation of the DMS shown in panel D
for increasing sequence divergence, 
both for experimental and in silico sequences (red curve and points).
Right: standard deviation of the $\Delta \Delta F$ of the same mutations shown in the left panel, with the same color code.
}
\label{fig:5}
\end{figure*}

To further validate our evolutionary model, we tested our results for contingency and entrenchment against a different set of experimental data. 
The authors of Ref.~\cite{park2022epistatic} reconstructed a phylogenetic lineage of a protein belonging to the DNA Binding Domain (DBD) family, from an ancient ancestor down to a present-day sequence, with a sequence divergence of about $35\%$ between the ancestor and the most recent sequence.
Then, they experimentally measured the effect of all mutations, i.e.~a DMS, around several protein sequences along the lineage, with variable sequence divergence. In line with expectations derived from our model, contingency and entrenchment were experimentally detected. We can then compare these experimental results with our in silico results by mapping the in silico $\Delta E$ to experimental $\Delta F$ using the non-linear mapping $\Delta F = \phi(\Delta E)$ shown in Fig.~\ref{fig:5}A (see SI, section 3, for the construction of the mapping from DMS data). Our model correctly reproduces the trends of contingency and entrenchment (SI, Fig.~S11), both in simulated trajectories or in experimental sequences scored with the DCA energy. Yet, the effect is very small as compared to the variability in the experimental measurements and our model is completely unable of predicting a single change of mutational effect $\Delta \Delta F$ (SI, Fig.~S12). This is explainable by the fact that our model parameters reflect a global fitness landscape learned on diverged homologs, hence it struggles at pointedly reproducing the fine scale effects caused by single mutations close to a wild type leading to contingency and entrenchment.

\section*{Epistasis in the Deep Mutational Scans of distant homologs}

Thanks to our model, we can obtain additional insight into how the effects of single mutations change in distant homologs.
In particular, we consider the DMS collecting the fitness variations of all single mutations around multiple reference sequences in the same family. We investigate how DMS correlate as a function of evolutionary distance between the reference sequences, both from our model and from experiments. In the absence of epistasis, the different DMS (for non-mutated sites) would be perfectly correlated because the same mutation has the same effect across all possible contexts. Hence, a decay of the correlation between DMS with increasing sequence divergence is a strong test for epistasis.

We consider as reference wild type (WT) the ancestral sequence reconstructed in~\cite{park2022epistatic}, and from the same experiments we retrieve its DMS and that of seven sequences along the evolutionary trajectory starting from the ancestral protein and reaching {\it C. teleta} SR, i.e.~the upper purple branch in Ref.~\cite[Fig.1B]{park2022epistatic}. 
For the evolutionary simulation, we start from the same WT and we run MCMC until reaching a sequence divergence of 10\%, 30\% and 60\%. The first two values (10\%, 30\%) are also reached in the experiment, but since our evolutionary trajectory is stochastic, the in silico sequences do not coincide with those of the experiment. The last value (60\%) is not reached in the experiment and corresponds, according to our model, to a divergence at which epistatic effects become strongly visible.

In Fig.~\ref{fig:5}B, we select the sites that mutated between the WT and each diverged sequence, both in the experiment and in our simulations. To assess the influence of the diverging sequence context on single mutational effects, we insert the WT residue into the diverged sequence, and the diverged sequence residue into the WT. 
The two corresponding in silico $\Delta E$ are mapped to experimental $\Delta F$ using again the non-linear mapping $\Delta F = \phi(\Delta E)$ shown in Fig.~\ref{fig:5}A (SI, section 3). In the absence of epistasis, the two reversion mutations should have exactly opposite effects, but due to epistatic interactions, we observe that, as the sequence context diverges, the effect of reverting mutations departs from the diagonal. In addition, we observe that as the sequences diverge from the wild type, the number of mutations in the shaded region of Fig.~\ref{fig:5}B (i.e. those that are deleterious in both contexts) grows. Following Ref.~\cite{chen2023understanding}, we call this type of residues ``2-wt entrenched''. To obtain a comprehensive summary of this phenomenon, we plot in Fig.~\ref{fig:5}C the fraction of 2-wt entrenched sites as a function of sequence divergence both in our simulations and in experiments, collecting every possible distance combination of the seven reconstructed proteins of Ref.~\cite{park2022epistatic}. 
We observe a good agreement between the results, and we confirm that our model can extend the range of observations beyond what is currently accessible to experiments.

We also address epistatic effects taking place on non-mutated residues. In Fig.~\ref{fig:5}D we compare the in silico DMS of the wild type and of the evolved sequence, restricted to the non-mutated residues, for the same three values of sequence divergence. From this plot, we clearly observe that mutational effects become less and less correlated upon increasing divergence of the context. To provide a more quantitative description of such epistatic effect, we compare the Spearman correlation of the DMSs on non-mutated sites shown in Fig.~\ref{fig:5}D, and (minus) the Spearman correlation of the mutation reversions presented in Fig.~\ref{fig:5}B; both are shown in Fig.~\ref{fig:5}E (left panel). 
Once again, we observe a very similar trend in experiments and simulations, confirming the validity of the model and highlighting a milder epistatic effect on non-mutated sites, as opposed to a stronger drop in correlation for mutation reversions.
In order to directly compare with the analysis of Ref.~\cite{park2022epistatic},  in Fig.~\ref{fig:5}E (right panel) we show
the standard deviation of the $\Delta \Delta F$ as a function of sequence divergence, both for the mutation reversions (Fig.~\ref{fig:5}B) and for identical mutations on non-mutated sites (Fig.~\ref{fig:5}D). We observe that such quantity increases as sequences diverge from each other, first linearly at small distances as shown in Ref.~\cite[Fig.2F]{park2022epistatic}, then in a more non-linear fashion when epistatic effects become more pronounced, i.e. at distances larger than $\approx 40\%$.

In summary, we have shown that epistasis emerges as a critical factor influencing the predictability of mutational effects in distant homologs, and that our model is able to reproduce such phenomenon coherently with what has been observed in recent experiments, confirming the reliability of our evolutionary simulations as a proxy for investigating protein evolution and complementing experimental results.

\section*{Discussion}

Our model offers a comprehensive framework to quantitatively investigate the dynamics of protein evolution, providing valuable insights bridging different time scales and evolutionary contexts from individual mutations to deep sequence divergence. Leveraging the established data-driven DCA approach based on Boltzmann Machines (bmDCA) to construct fitness landscapes, 
and complementing it with an appropriate
Monte Carlo Markov Chain (MCMC) dynamics comprising single-nucleotide mutations as well as indels (Fig.~\ref{fig:1}A),
we constructed a framework capable of accurately reproducing statistical features observed in protein sequences both in experimental and natural evolution. We were able to unveil the interplay between mutations and selection pressures, offering a deep understanding of evolutionary trajectories spanning from the short-term in vitro evolution (Fig.~\ref{fig:1}B) to long-term natural evolutionary scales (Fig.~\ref{fig:1}C-D).

The primary contribution of our work lies in the identification and characterization of emergent long epistatic-driven evolutionary timescales. We showed how mutations accumulate and interact within the sequence landscape, influencing the trajectory of protein evolution. The emergent collective epistatic behavior takes place way beyond the short time scales accessed experimentally, but much before the ones reached by natural evolution. Our findings not only challenge independent site evolutionary models by emphasizing the importance of epistatic effects, but also provide quantitative insights into the mechanisms governing evolutionary dynamics. In fact, we are able (Fig.~\ref{fig:2}A-B) to predict the residues undergoing epistatic-driven evolution just by using the natural data and our inferred model. 

Furthermore, our investigations shed light on the phenomena of contingency and entrenchment, which play pivotal roles in shaping evolutionary paths. We quantitatively assessed how mutations are contingent upon previous genetic changes and how they can become entrenched due to subsequent mutations (Fig.~\ref{fig:3}), thereby offering a perspective on the dynamics of adaptation. 
Our results were tested against experiments to validate the reliability of our evolutionary simulations exploiting a non linear mapping between experimental fitness and model energy (SI, Fig.~S11). Our computational predictions capture the qualitative trend of the empirical observations of contingency and entrenchment. 

Importantly, we also investigated how the collective epistatic behavior of residues plays a role in DMSs of distant homologs (Fig.~\ref{fig:5}). Correlations between DMSs decrease with sequence divergence and mutation reversions get increasingly more deleterious showing pervasive entrenchment of residues, in line with the results of Ref.~\cite{chen2023understanding}. Such an increasing variability of differences between the same mutation at different times suggest the key role of epistasis in limiting predictability. Even in this case, the simulations results are coherent with recent experimental tests.

Overall, our work contributes to a deeper understanding of protein evolution by elucidating the complex interplay between mutations, selection pressures, and evolutionary timescales. By providing quantitative insights and validation against experimental data, we offer a comprehensive framework for exploring the mechanisms driving protein diversity and adaptation. We believe that our model can be used in future research to optimize experimental protocols for in vitro evolution, e.g.~by introducing a time-dependent selection pressure. The collective nature of epistasis, i.e.~the dynamical correlations between epistatic mutational events, could also be better understood by a deeper investigation of the model-generated trajectories, and tested in future experiments.

\acknowledgments

We thank Thierry Mora, Rama Ranganathan, Juan Rodriguez-Rivas, Joe Thornton and Aleksandra Walczak for many useful discussions.

\section*{Methods}
\subsection*{Sequence and Deep mutational scan data}

We include in our analysis the $\beta$-lactamase sequence data coming from
the in vitro evolution experiments  
on TEM-1~\cite{fantini2020protein} and PSE-1~\cite{stiffler2020protein}. 
For Ref.~\cite{fantini2020protein},
sequences with more than six gaps were discarded
as a quality control. 
The resulting MSAs of experimentally evolved sequences
have 202 sites and 456,871 sequences for PSE-1 (round 20),
and 34,431 sequences for TEM-1 (generation 12). The natural MSA alignments were generated running the {\tt hmmsearch} command from the HMMer software suite on the
UniProt database. Insertions
were removed, and sequences with more than $10\%$ gaps and
duplicated sequences were excluded. Any sequence closer than $80\%$ to the wild types
TEM-1, PSE-1 was excluded from the alignment. 

For the DBD family,
we retrieve from Ref.~\cite{park2022epistatic} the DMS of seven sequences along the evolutionary trajectory starting from the ancestral DBD protein and reaching {\it C. teleta} SR, i.e. the upper purple branch in Ref.~\cite[Fig.1B]{park2022epistatic}. 
To construct the MSA of natural sequences,
we used as seed the alignment from Ref.~\cite{park2022epistatic} 
(76 sites and 221 sequences) and ran 
{\tt hmmsearch} on {\tt uniref90}, 
excluding
sequences with more than 20\% gaps.
Any sequence closer than 80\% to any of the 9 sequences tested in Ref.~\cite{park2022epistatic} was excluded from the alignment.

While the choice of removing from the natural alignments
the sequences that are close to the reference wild types (TEM-1, PSE-1 for the $\beta$-lactamase family and the 9 sequences tested in Ref.~\cite{park2022epistatic} for the DBD family) does not strongly influence the results, it was done for fairness of comparison, in such a way that the
bmDCA learning is not explicitly informed on the local fitness landscape by natural sequences in the vicinity of the wild types of interest. \\ In addition, to account for similar sequences in datasets, i.e. to reduce the bias in the training data, each sequence is weighted based on how similar it is to other sequences. For this purpose, we used the standard procedure of DCA, as detailed in Ref.~\cite{morcos2011direct}. The re-weighting procedure is important also when performing principal component analysis. In Fig.~\ref{fig:1}D the principal components of natural sequences have been computed taking into account the weight of each sequence. Moreover, only a subset of natural sequences (sampled according to the same weights introduced before and of the same size as the in silico library) has been used to realize the plots. The same holds true for every other plot where we compare sampled sequences with natural sequences.

\subsection*{Statistical inference procedure}
The statistical tool used to infer the protein fitness landscape for the generative model is known as Boltzmann Machine Learning (bmDCA)  and it is currently one of the most widely used techniques in the context of DCA. The key assumption of this inverse statistical physics technique is to admit that proteins of the same family (i.e. the rows of the MSA) can be considered as (not necessarily i.i.d.) samples from a probability distribution that takes the following Boltzmann-like expression:
\begin{equation}
    P(a_1,...,a_L)=\frac{e^{-E(a_1,...,a_L)}}{Z}
\end{equation}
where $a_i$ is the amino acid in the i-th position of the sequence $\mathbf{a}=(a_1,\cdots,a_L)$ of length $L$, and $Z=\sum_{\mathbf{a}} e^{-E(\mathbf{a})}$ is the partition function commonly encountered in statistical physics. In this sense, the energy provides a high-dimensional landscape in which non-functional sequences correspond to high-energy peaks, whereas functional sequences populate the valleys. Following the Maximum Entropy Principle (MEP), we choose a set of observables $O_{\alpha}(\mathbf{a})$ of the data set that we want to reproduce with our probabilistic model. These will act as constraints on our functional entropy maximization.
In bmDCA one selects the one-point frequencies $p_i(a)=\langle \delta_{a,a_i} \rangle_{P}$ and
 the two-point frequencies $p_{ij}(a,b)=\langle \delta_{a,a_i}\delta_{b,a_j}\rangle_{P}$
for every possible $i,j,a,b$.
Maximizing the entropy is equal to finding the maximum of the data likelihood in a Bayesian context~\cite{cocco2018inverse}.
For our study, 
the natural MSA of the $\beta$-lactamase and DBD families, constructed as indicated above, were used to train a Potts model
using bmDCA~\cite{figliuzzi2018pairwise} in the implementation of Ref.~\cite{muntoni2021adabmdca} with standard learning parameters.

\subsection*{Evolutionary model}
Given the learned protein fitness landscape, we simulate evolution as a stochastic process over such landscape. To correctly reproduce how mutations take place during evolution, we introduce for each site $i$ the corresponding three-nucleotides codon $c_i$, and design an evolutionary model on the nucleotide sequence $\textbf{c} = (c_1,\cdots, c_L)$, which is then reflected at the amino acid level $\mathbf{a}(\mathbf{c})$ using the genetic code. 

Let us denote by $f(c_i|a_i)$ the frequency with which codon $c_i$ codes for amino acid $a_i$. The simplest choice we use in this work is uniform, $f(c_i|a_i) = 1/N(a_i)$ where $N(a_i)$ is the number of codons that code for amino acid $a_i$.
One can consider different choices, e.g. by estimating $f(c_i|a_i)$ from the codon bias of the species in which a given experiment is performed. Here we just check that using different codon bias as input produces the expected codon usages in the simulated sequences (SI, Fig.~S9). However, for long-term evolution it does not make sense to consider the species-dependent codon bias, as natural evolution takes place across different species. 
From the relation
\begin{equation}
P(\mathbf{c}) = P[\mathbf{a}(\mathbf{c})] \prod_i f(c_i|a_i) \ ,
\end{equation}
we can introduce a novel nucleotide sequence energy 
\begin{equation}
    E(\mathbf{c}) = - T \log P(\mathbf{c}) = E[\mathbf{a}(\mathbf{c})] - T \sum_{i=1}^L \log f(c_i|a_i) \ ,
\end{equation}
which we use in our evolutionary dynamics.

In order to realistically mimic protein evolution, we introduce in our stochastic dynamics three different evolutionary phenomena:
\begin{itemize}
    \item single-nucleotide mutations $\to$ one nucleotide is replaced by another;
    \item deletions $\to$ an entire codon is replaced by a gap (3 nucleotides get deleted);
    \item insertions $\to$ a gap is replaced by an entire codon (3 nucleotides get inserted).
\end{itemize}
Here, gaps are positions in a protein sequence where one amino acid is missing.
We choose to work with Gibbs and Metropolis sampling, combined with Markov Chain Monte Carlo (MCMC). For our purposes, we develop a mixed sampler that proceeds in the following way:
\begin{itemize}
    \item start with a nucleotide sequence;
    \item with probability $p$ do a Metropolis move modeling insertion/deletion of a codon;
    \item with probability $1-p$ do a Gibbs move modeling single-nucleotide mutations;
    \item iterate the process with the updated sequence as input.
\end{itemize}

The Gibbs move proceeds in the following way:
\begin{itemize}
\item choose a nucleotide position $i$ at random (considering non-gapped positions only)
\item sample the new nucleotide $n'_{i}$ according to the conditional probability $P(n'_{i}|\mathbf{n}_{\setminus i})$
of all possible single-nucleotide mutations in that position.
\end{itemize}
The Metropolis move works instead on the codon space in the following way:
\begin{itemize}
\item Choose a site $i$ at random;
\item Propose a move $c_i \to c_i'$ to a new state through a proposal matrix $\Omega(c_i \to c_i')$;
\item Accept the move according to the probability $p(c_i \to c_i')$.
\end{itemize}
The probability of changing codon is
\begin{equation}\begin{split}
p(c_i \to c_i') &= \min\left(1,  e^{-[ E(\mathbf{c}') - E(\mathbf{c}) ]/T} \right) = \\
&= \min\left(1, \frac{f(c_i'|a_i')e^{-E[\mathbf{a}(\mathbf{c}')]/T}}{f(c_i|a_i)e^{-E[\mathbf{a}(\mathbf{c})]/T}}   \right) \ , 
\end{split}\end{equation}
and the proposal matrix is the 
same on all sites,
\[
\Omega(c \to c') =
\begin{cases}
    \alpha, & \text{if } c,c' = G  \\
    \beta, & \text{if } c = G, c' \in C \text{ or } c \in C, c' =  G \\
    \gamma, & \text{if } c = c' \in C \\ 0, & \text{if } c \in C \text{ and } c' \in C \text{ with } c \neq c'
\end{cases}.
\] Here, $G$ is the gap codon and $C$ is the set of 64 possible codons observed in nature. This matrix has been constructed in order to admit insertion and deletions, disallow substitutions of codons ($c \in C \to c' \in C$), while still being normalized and symmetric. From an efficiency standpoint, one could argue that the non-null diagonal makes the algorithm quite slow since it proposes useless moves that do not change the sequences. Unfortunately $\alpha$ and $\gamma$ are necessary for normalization, still they can be reduced by maximizing insertion/deletion proposals choosing $\beta=\frac{1}{64}$, which consequently imposes $\gamma=\frac{63}{64}$ and $\alpha=0$. As a result of this procedure, both Metropolis and Gibbs moves enforce detailed balance, hence the entire algorithm is guaranteed to converge to the correct probability distribution of natural sequences (see SI).

\subsection*{Data and Code Availability}
The data and code utilized in this study are available to facilitate reproducibility and further research. The code implemented for analysis and modeling, the notebooks to reproduce figures and the utilized data can be accessed through the following repository \url{https://github.com/leonardodibari/Gen.jl/tree/main}.

\newcommand{\beginsupplement}{
        \clearpage
        \onecolumngrid
        \setcounter{section}{0}
        \renewcommand{\thesection}{S\arabic{section}}
        \setcounter{equation}{0}
        \renewcommand{\theequation}{S\arabic{equation}}
        \setcounter{table}{0}
        \renewcommand{\thetable}{S\arabic{table}}
        \setcounter{figure}{0}
        \renewcommand{\thefigure}{S\arabic{figure}}
     }
\beginsupplement

\section*{Supplementary Information}

\section{Evolutionary Algorithm}

Sampling from high-dimensional probability distributions usually requires advanced algorithms, such as Markov Chain Monte Carlo (MCMC) sampling. For spin models defined over categorical variables, like Potts models, many sampling algorithms are available. For example, Metropolis-Hastings and Gibbs Monte Carlo are both utilized as a sampling technique in the implementation of Boltzmann learning to infer protein sequence landscapes~\cite{muntoni2021adabmdca}. 
In this context, MCMC sampling generates equilibrium sequences via long simulations, ensuring the proper exploration of sequence space. However, MCMC sampling can also be used to model the dynamics of biological systems by properly interpreting the Markov chain configurations~\cite{de2020epistatic}. In particular, it was shown that Markov chain sampling can be used to model short-term protein evolution in neutral genetic drift experiments: short simulations performed with Gibbs sampling, with the addition of the constraints of the genetic code, accurately reproduced many statistical features of the experimental data~\cite{bisardi2022modeling}.
In the next sections, we discuss how to extend the approach of Refs.~\cite{de2020epistatic,bisardi2022modeling} to sample long chains that reach equilibrium, thereby modeling protein evolution over billions of years, while also accurately incorporating the amino acid accessibility dictated by the genetic code. 

\subsection{Model definition over nucleotide space}
As a generative model of amino acid sequences, we consider a Potts model
\begin{equation}
  P({\bf{a}}) = \frac{  e^{\ -\beta H ({\bf{a}})} }{Z  }
\label{eq2:p_amino}
\end{equation}
whose parameters of the Hamiltonian $H$ have been inferred from the natural sequences of a protein family and $\beta = 1/T$ is the inverse temperature. Using this model, we aim to define an evolutionary dynamics taking into account that evolution happens in nucleotide space. This necessitates specifying the probability distribution of $P({\bf{a}})$ over nucleotide sequences instead of amino acid sequences. A straightforward procedure to accomplish this consists of uniformly distributing the probability associated with an amino acid sequence across all its synonymous nucleotide sequences.
To formalize this concept, we define an amino acid sequence of length $L$ as ${\bf{a}} = (a_1, a_2, ..., a_L)$ and a corresponding nucleotide sequence as ${\bf{n}} = (n_{1_1}, n_{1_2}, n_{1_3}, n_{2_1},  ..., n_{L_3})$. 
The function $A(\mathbf{n})$ applies the genetic code to transform the nucleotide sequence $\mathbf{n}$ into its respective amino acid sequence $\mathbf{a}$. For each amino acid sequence ${\bf{a}}$, we also define the set 
\begin{equation}
    \mathbf{\Gamma}(\mathbf{a}) = \{ {\bf{n}} : A({\bf{n}}) = {\bf{a}}  \},
\end{equation}
and the function $N(\mathbf{a})$ =  $|\mathbf{\Gamma}(\mathbf{a})|$, i.e. the cardinality of $\mathbf{\Gamma}(\mathbf{a})$, which counts the number of synonymous nucleotide sequences coding for a given amino acid sequence ${\bf{a}}$. Because each amino acid is coded independently, $N({\bf{a}})$ factorizes over the amino acids and can be rewritten as:
\begin{equation}
    N(\mathbf{a}) = \prod^L_{i = 1} N(a_i)
\end{equation}
where $N(a)$ is the number of synonymous codons coding for amino acid $a$. We now have all the elements to define the probability distribution of a new model $\mathcal{P}({\bf{n}})$ over nucleotide space: 
\begin{equation}
    \mathcal{P}({\bf{n}}) = \frac{P(A({\bf{n}}))}{N(A({\bf{n}}))} = \frac{1}{N(A({\bf{n}}))}\frac{  e^{\ -\beta
H ( A({\bf{n}}))} }{Z  } = \frac{  e^{\ -\beta \mathcal{H} ({\bf{n}})} }{Z  }
\label{eq2:p_nucleo}
\end{equation}
such that the new Hamiltonian is defined as:
\begin{equation}
\begin{aligned}
\mathcal{H}({\bf{n}}) &= H(A({\bf{n}})) + T \log{[N(A({\bf{n}}))]} \\ &= H(\mathbf{a}) + T \sum^L_{i = 1} \log{[N(a_i)]}.
\end{aligned}
\label{eq2:new-hamiltonian}
\end{equation}
The Hamiltonian in \eqref{eq2:new-hamiltonian} incorporates a novel term compared to the standard Potts Hamiltonian defined over amino acid space. Namely, 
\begin{equation}
    T \sum^L_{i = 1} \log{[N(a_i)]}
\label{eq2:t-factor}
\end{equation}
which accounts for the entropic contribution arising from the degeneracy of synonymous DNA sequences. 

The term in \eqref{eq2:t-factor} introduces a correction that disfavors nucleotide sequences with high genetic redundancy by assigning them higher energy, i.e. lower probability, compared to sequences with low degeneracy which are assigned lower energy, i.e. higher probability. Consequently, the original Potts probability of amino acid sequence $\mathbf{a}$ defined in \eqref{eq2:p_amino} is correctly recovered after summing over all synonymous nucleotide sequences:
\begin{equation}
    P(\mathbf{a}) = \sum_{\mathbf{n} \in \mathbf{\Gamma}(\mathbf{a})} \frac{  e^{\ -\beta H(A(\mathbf{n})) } }{Z  }.
\end{equation}

An alternative approach to define the probability distribution on nucleotide space is to introduce a codon bias, i.e. to assign non-uniform weights to the codons, accounting for the specific codon usages of different species. Let us alternatively denote a nucleotide sequence by a sequence of codons $\mathbf{c} = (c_1,\cdots, c_L)$, which is then reflected at the amino acid level $\mathbf{a}(\mathbf{c})$, using the genetic code. We define also $f(c_i|a_i)$ the frequency with which codon $c_i$ codes for amino acid $a_i$. 
From the relation
\begin{equation}
P(\mathbf{c}) = P[\mathbf{a}(\mathbf{c})] \prod_i f(c_i|a_i) \ ,
\end{equation}
we can introduce a novel nucleotide sequence energy 
\begin{equation}
    \mathcal{H}(\mathbf{c}) = - T \log P(\mathbf{c}) = E[\mathbf{a}(\mathbf{c})] - T \sum_{i=1}^L \log f(c_i|a_i) \ ,
\end{equation} 

If  $f(c_i|a_i) = 1/N(a_i)$ we fall back in the previous description without codon bias, but one can consider different choices, e.g. by estimating $f(c_i|a_i)$ from the codon bias of the species in which a given experiment is performed. However, for long term evolution, it does not make sense to consider the species-dependent codon bias, as natural evolution takes place across different species. 
In Fig.~S9 we show the codon usage measured from in silico sequences generated via simulations with or without codon bias.
In our case, introducing a codon bias for the experiments at short time scales is not relevant for modeling purposes, because the mutational process was performed in vitro, hence we assume no codon bias to carry over our analysis.

\subsection{Description of the algorithm}

Detailed balance (DB) is a mathematical condition ensuring that the configurations generated by a Markov process converge to a stationary distribution. In our case, DB ensures that the Markov chains converge to the statistics of natural amino acid sequences that we used to infer the Potts model. Unfortunately, the sampling dynamics on nucleotides introduced in Ref.~\cite{bisardi2022modeling} does not respect the DB condition. Adapting the sampling algorithm on nucleotides to satisfy DB is the goal of this section. DB is a statement about the transition probabilities $\pi$ between configurations of the Markov Chain. In the case of nucleotide sequences, for every pair of sequences $\mathbf{n}, \mathbf{n}'$, the following expression must hold:
\begin{equation}
\mathcal{P}({\bf{n}})\pi({\bf{n}} \rightarrow {\bf{n}}')
 =\mathcal{P}({\bf{n}}') \pi({\bf{n}}' \rightarrow {\bf{n}}) \ .
\end{equation}
The condition requires that, at equilibrium, the probability flux  $\mathbf{n} \rightarrow \mathbf{n}'$ equals the one from $\mathbf{n}' \rightarrow \mathbf{n}$, i.e. the probabilities of $\mathbf{n}$ and $\mathbf{n}'$ do not change. A vast class of Markov Chain Monte Carlo algorithms depends on finding a suitable transition matrix that satisfies this equation. In addition to that, we want $\pi$ to model a plausible evolutionary dynamics on nucleotides. To this aim,  we need to account for at least three processes:
\begin{itemize}
    \item single-nucleotide mutations: a nucleotide is replaced by another one
    \item amino acid deletion: a codon is lost from the sequence, removing 3 nucleotides
    \item amino acid insertion: a codon is inserted in the sequence, adding 3 nucleotides
\end{itemize}
In terms of indels, we model only triplets on nucleotides because inserting or deleting single (or pairs of) bases results in a frameshift, misaligning subsequent codons in the reading frame and leading with very high probability to non-functional sequences. On top of this, in terms of the amino acid sequence, this move is highly non-local, thus hard to model. As a consequence, we assume that this process has zero probability and we exclude it from the dynamics. While this approach represents a simplification of the actual indel statistics, we contend that modeling triplets of indel maintains a biological foundation. Replication slippage, one of the most well-understood mechanisms generating indels in protein evolution, exemplifies this biological basis. During DNA replication, the DNA polymerase sometimes pauses and disconnects from the DNA. This pause can allow the new, growing DNA strand to momentarily detach and then accidentally reconnect to a similar sequence either ahead or behind the original spot. When the DNA polymerase restarts copying, it may either miss or repeat a section of the DNA, creating a deletion or an insertion in the new DNA strand. 

Including both indel mutations and single-nucleotide substitutions in a single MCMC framework is not trivial, since they act differently on variables and satisfaction of DB cannot be guaranteed. To overcome this issue, we suggest an approach that combines Gibbs and Metropolis sampling. We devised a mixed sampler operating as follows:
\begin{itemize}
    \item with probability $p$, execute a Metropolis move simulating codon indels, i.e. operating on triplets of nucleotides.
    \item with probability $1-p$, perform a Gibbs move simulating a single-nucleotide mutation.
\end{itemize}
Note that this also reflects the different nature of the two processes: our dynamics determines first which process happens, and then its outcome. 

\subsubsection{Indels: Metropolis sampling}
To model codon insertions and deletions (indels), we introduce a new symbol: the triple gap "-\,-\,-" which we denote with $c^0$. This codon, equivalent to an amino acid gap, models the deletion of a codon within a nucleotide sequence. We also designate with \(c^k\), \(k \in \{1, ..., 64 \}\), all other amino-acid-coding codons, including the stop codons. 

We can now we decompose the Metropolis transition matrix \(\pi({\bf{n}} \rightarrow {\bf{n}}')\) into two components: a proposal term \(p({\bf{n}} \rightarrow {\bf{n}}')\) and an acceptance term \(\alpha({\bf{n}} \rightarrow {\bf{n}}')\). 
Since we exclusively focus on single codon substitutions, we only need to consider transitions \( \mathbf{n} \rightarrow \mathbf{n}' \) such that in codon language \( \mathbf{n} = \mathbf{c} = (c_1, c_2, \ldots, c_i, \ldots, c_L) \) and \( \mathbf{n}'=\mathbf{c}' = (c_1, c_2, \ldots, c_i', \ldots, c_L) \). As a consequence, we can focus on the single-codon transition matrices \(p(c \rightarrow c')\) and \(\alpha(c \rightarrow c')\).

We restrict our proposal to mutations from amino-acid-coding codons towards the gap codon and vice versa, leading to the following proposal matrix \(p( c \rightarrow c')\) where the missing elements are set to zero:
\begin{equation}
\begin{array}{c|ccccc}
& c^0 & c^1 & c^2 & \cdots & c^{64} \\
\hline
c^0 & \eta & \beta & \beta & \cdots & \beta \\
c^1 & \beta & \gamma & & 0 & \\
c^2 & \beta & & \gamma &  &  \\
\vdots & \vdots & 0 & & \ddots &  \\
c^{64} & \beta & & & & \gamma \\
\end{array}
\label{eq2:proposal-metro}
\end{equation}

This symmetric matrix accommodates insertions and deletions via the parameter $\beta$ and prohibits amino acid substitutions:
\begin{equation}
    p( c^m \rightarrow c^n) = 0 \iff m \neq n \quad \text{or} \quad m,n > 0.
\end{equation} 
The parameters $\eta$ and $\gamma$ are used to normalize the proposal probability and correspond to “empty” moves not changing the state of the codon. To speed up the algorithm, we want to maximize $\beta$ while maintaining the proposal probability normalized. As a result we get $\eta = 0$,  $\beta = 1/64$ and $\gamma = 63/64$. This means that when we select an amino-acid-coding codon, in $63$ out of $64$ attempts we propose the same codon, i.e. nothing changes, and only in one case we do propose the triple gap. If a stop codon is proposed, it never gets accepted.

We are now in a position to describe how the Metropolis sampling algorithm for indels works. For each Monte Carlo step:

\begin{enumerate}
    \item A sequence position \(i \in \{1, \ldots, L\}\) is randomly selected, along with its corresponding $c_i$ codon.
    \item A new codon $c_i'$ is proposed though the proposal matrix ${p(c_i \rightarrow c_i')}$
    \item The acceptance probability of the proposed transition $c_i \rightarrow c_i'$ is computed according to the Metropolis prescription:
    \begin{equation}
    \alpha(c \rightarrow c') = min\left(1, \frac{\mathcal{P}({\bf{n}}')}{\mathcal{P}({\bf{n}})}\right) = min \left( 1, e^{ -\beta \left[ \mathcal{H}(\mathbf{n}') - \mathcal{H}(\mathbf{n}) \right] }  \right)
\label{eq2:acc-metropolis}
\end{equation}
    \item Codon $c_i'$ is accepted with probability $\alpha(c_i \rightarrow c_i')$ or refused with probability $1 - \alpha(c_i \rightarrow c_i')$.
\end{enumerate}

Thanks to the fact that the proposal matrix in \eqref{eq2:proposal-metro} is symmetric and that we accept codon mutations according to \eqref{eq2:acc-metropolis}, detailed balance is guaranteed. 

We highlight that insertions and deletions are a crucial feature of the algorithm to converge to the correct stationary probability distribution and not just a detail that has been introduced for completeness. Despite having a lower statistical energy, sequences sampled without indels do not correctly cover the main two principal directions of natural data in PCA space and they have a lower diversity/pairwise hamming with respect 
 to natural data (Fig.~S10).

\subsubsection{Point mutations: Gibbs sampling }

The second type of mutation that we need to consider is single-nucleotide substitutions. We resort to Gibbs Monte Carlo sampling as in Ref.~\cite{bisardi2022modeling}, but we
generalize that approach by introducing DB for accurate long-term sampling. Gibbs sampling works by iteratively emitting new variables from their conditional distribution, given the current value of the rest of the sequence. According to \eqref{eq2:p_nucleo}, we can express the conditional probability of $n_{i_{k}}$ (i.e. nucleotide $n$ belonging to the $i$-th codon, position $k$) given the rest of the sequence as:
\begin{equation}
\begin{aligned}
\mathcal{P}(n_{i_k}| \mathbf{n}_{-i_k}) &= \frac{\mathcal{P}(\mathbf{n})}{\mathcal{P}(\mathbf{n}_{-i_k})} = \frac{\mathcal{P}(\mathbf{n})}{\sum\limits_{n_{i_k}\!\in \mathcal{N}}\!\mathcal{P}(\mathbf{n})} \\
& =\frac{e^{-\beta 
\mathcal{H}(n_{1_{1}},\, n_{1_{2}}, \,\ldots, \,n_{i_{k}},\, \ldots,\, n_{L_{3}})}}
{\sum\limits_{n \in \mathcal{N}} e^{-\beta 
\mathcal{H}(n_{1_{1}},\, n_{1_{2}}, \,\ldots, \,n_{i_{k}}=n,\, \ldots,\, n_{L_{3}})}} 
\end{aligned}
\label{eq2:cond-gibbs}
\end{equation}
where $\mathcal{N} =$ \{A, C, G, T\}.

We can now present how a step of Gibbs Monte Carlo sampling works in our algorithm:
\begin{enumerate}
    \item A nucleotide position $i_k$, $i \in \{1, \ldots, L \} $ and $k \in \{ 1, 2, 3 \} $, is randomly selected, excluding currently gapped positions.
    \item In order to explicitly compute the value of \eqref{eq2:cond-gibbs}, the set of amino acids
    $\mathcal{A}_{i_k}$ is generated, which is the translation of the $4$ possible codons in position $i$ derived by inserting each of the nucleotides of $\mathcal{N}$ in position $i_k$.
    \item A new amino acid $n'$, and the respective amino acid $a'$, is sampled from the following probability distribution, computed exploiting the form of the Hamiltonian in 
    \eqref{eq2:new-hamiltonian}:
    \begin{equation}
    \begin{aligned}
    \mathcal{P}(n_{i_k} = n'| \mathbf{n}_{-i_k})  
    &=\frac{e^{-\beta 
    H(a_1,\,\ldots, \,a_i=a',\, \dots,\, a_L) - \log[N(a_1,\,\ldots, \,a_i=a',\, \dots,\, a_L)]}}
    {\sum\limits_{a \in \mathcal{A}_{i_{k}}} e^{-\beta 
    H(a_1,\,\ldots, \,a_i = a,\, \dots,\, a_L) - \log[N(a_1,\,\ldots, \,a_i=a,\, \dots,\, a_L)]}} 
    \\
    &= \frac{e^{\, \beta h_i(a') + \beta \sum_j J_{ij}(a', a_j) - \log[N(a')]}} { \sum\limits_{a \in \mathcal{A}_{i_{k}}}e^{\,\beta h_i(a) + \beta \sum_j J_{ij}(a, a_j) - \log[N(a)]}}
    \end{aligned}
    \label{eq2:conditional-amino}
    \end{equation}
    \item $n'$ is accepted unless it produces a stop codon. In either case, the procedure is restarted from point 1.
\end{enumerate}
\eqref{eq2:conditional-amino} is quick to evaluate, thanks to the many simplifications happening.  In particular, the fields $h$ and the couplings $J$ between unmutated sites, as well as the degeneracies $N(\mathbf{n})$ of unmutated codons appear in both the numerator and the denominator, so they can be divided out from the expression, leaving only parameters coupled to the mutated site.

\subsection{Comparison with previous algorithms}

The evolutionary model that has been thoroughly explained in the previous sections was developed by iteratively taking information from previous models. In fact, a first important contribution highlighted how Gibbs sampling Monte Carlo dynamics on epistatic landscape models~\cite{de2020epistatic} reproduces several statistical properties of protein evolution, while being consistent with neutral evolution theory, unifying observations from previously distinct evolutionary sequence models. A second contribution has applied this idea to modeling experimental protein evolution, which can only reach up to the short-term scale of $\simeq 20\%$ of divergence from the wildtype~\cite{fantini2020protein, stiffler2020protein}. It was shown that the inclusion of genetic code constraints on mutations leads to higher correlations with experimental data~\cite{bisardi2022modeling}. Despite these advances, the results of Ref.~\cite{bisardi2022modeling} were confined to (i) short timescales and (ii) sequences that do not contain gaps. 
In particular, because the selection procedure of Ref.~\cite{bisardi2022modeling} did not satisfy detailed balance,
the resulting dynamical model was not suited to reproduce the long-term equilibrium properties of protein family alignments. To improve upon previous results~\cite{de2020epistatic,bisardi2022modeling,alvarez2024vivo}, our new in silico evolutionary dynamics uses a simple, biologically motivated stochastic process that takes into account mutational constraints due 
to the genetic code, and the possibility of insertions and deletions, while still satisfying detailed balance. 

\clearpage

\section{Supplementary figures}

\subsection{Additional data on $\beta$-lactamases}

\begin{figure}[htbp]
\centering
\includegraphics[width=0.7\linewidth]{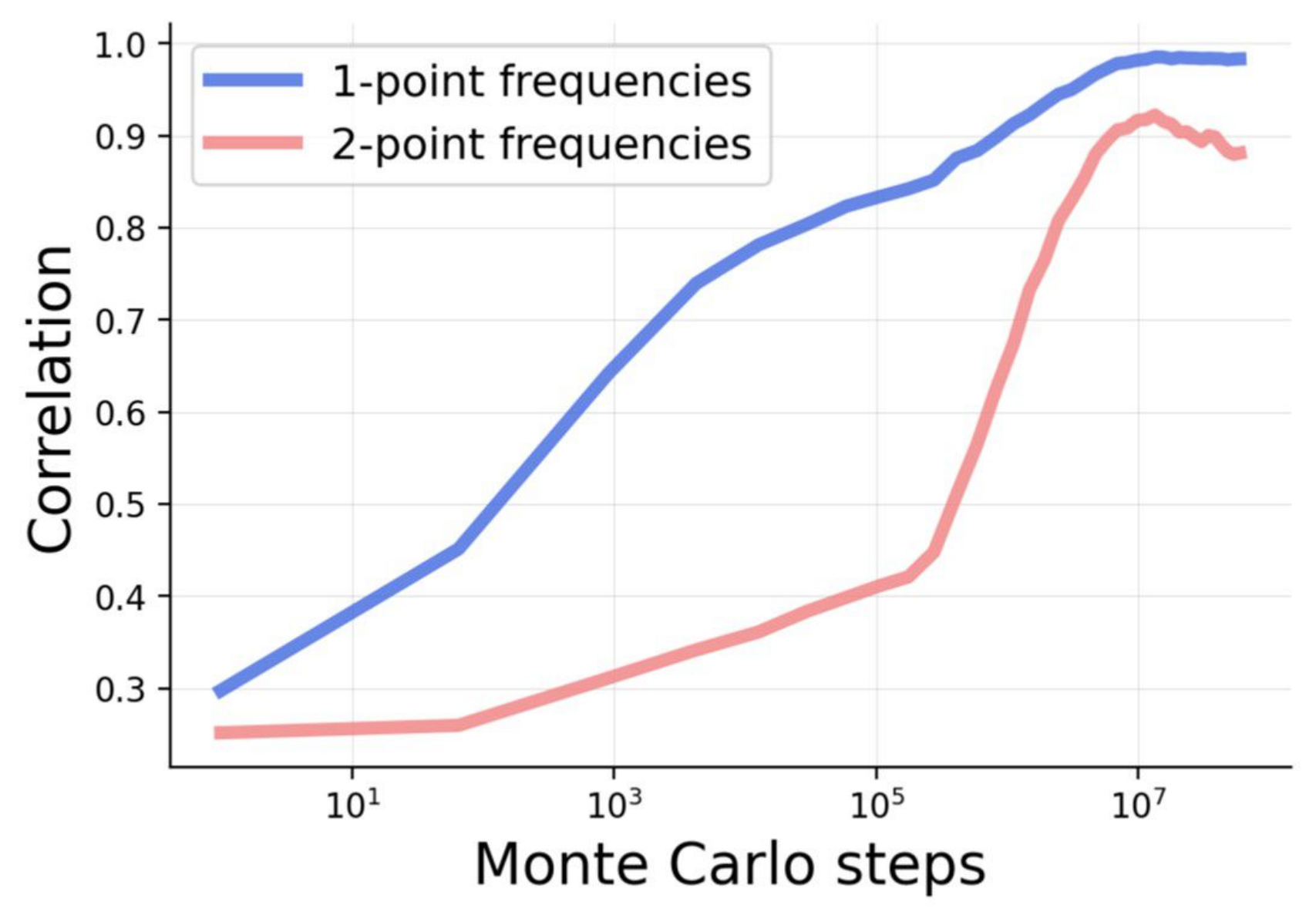}
\caption{{\bf Statistical properties of 1000 in silico sequences for the $\beta$-lactamase family at different times.}
Pearson correlation of the single-point $f_i(a)$ and connected two-point $c_{ij}(a,b)$ statistics between natural sequences
and a set of 1000 in silico sequences sampled over time.
The starting MSA is entirely composed of PSE-1 wt sequences. The plot shows how the independent Monte Carlo chains progressively diversify,
asymptotically reproducing the statistics of natural sequences.
}
\label{fig:s1}
\end{figure}

\clearpage
\subsection{Validation of the model for the DBD family}

\begin{figure}[htbp]
\centering
\includegraphics[width=.8\linewidth]{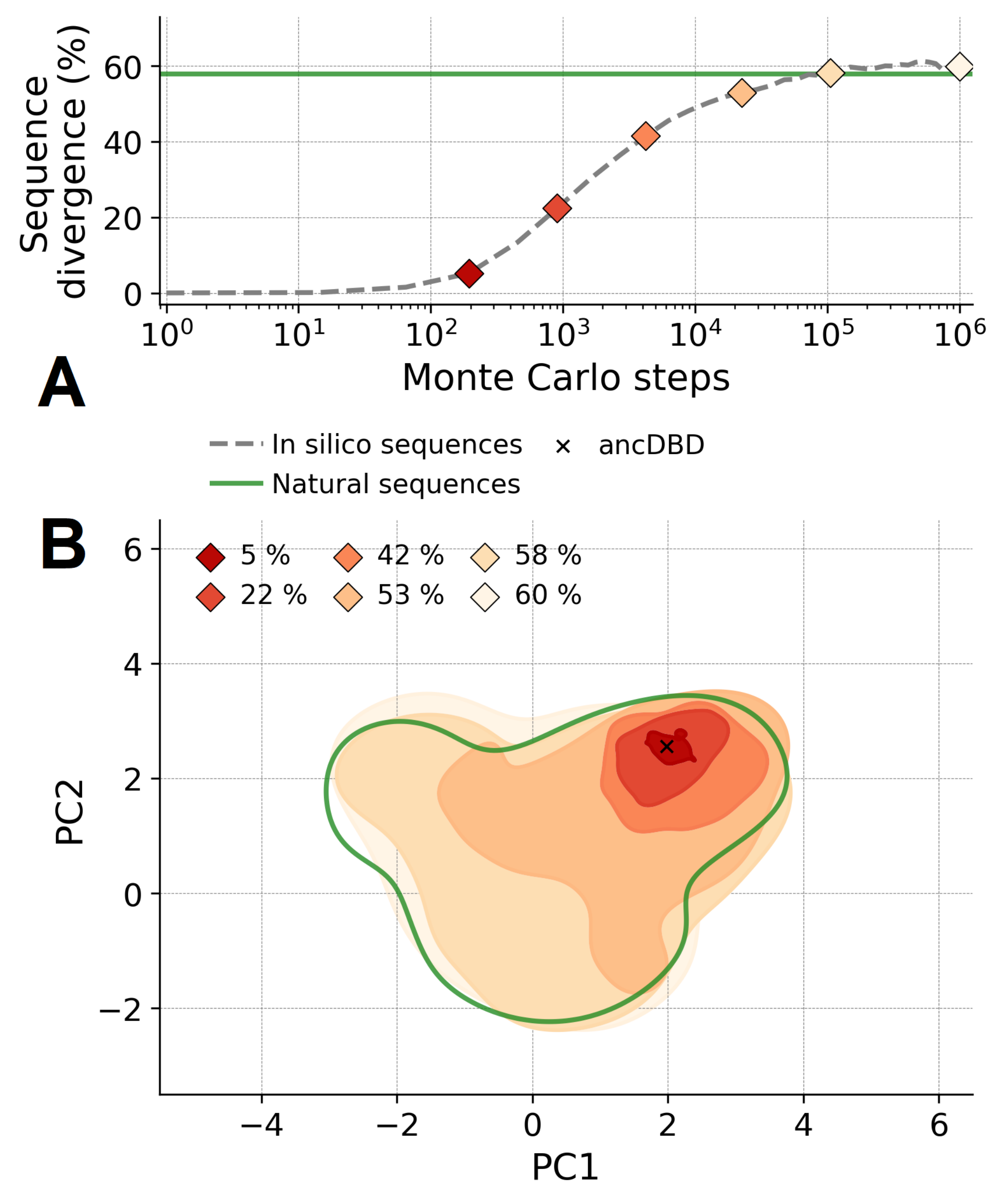}
\caption{{\bf Illustration and validation of the evolutionary model on the DBD protein family.}
(A) Average sequence divergence (normalized Hamming distance) from the reconstructed ancestral sequence in which simulation has started over 100 Markov chain realizations of the mutational dynamics at $\beta=1$. The green line depicts the average sequence divergence of the family of natural proteins used as a training set. (B) Projection over the two principal components (PCA) of the natural sequences (green) and
of the sequences sampled after a given number of in silico evolutionary steps (red to light orange, labeled by the average sequence divergence). }
\label{fig:s2}
\end{figure}

\begin{figure}[htbp]
\centering
\includegraphics[width=1.\linewidth]{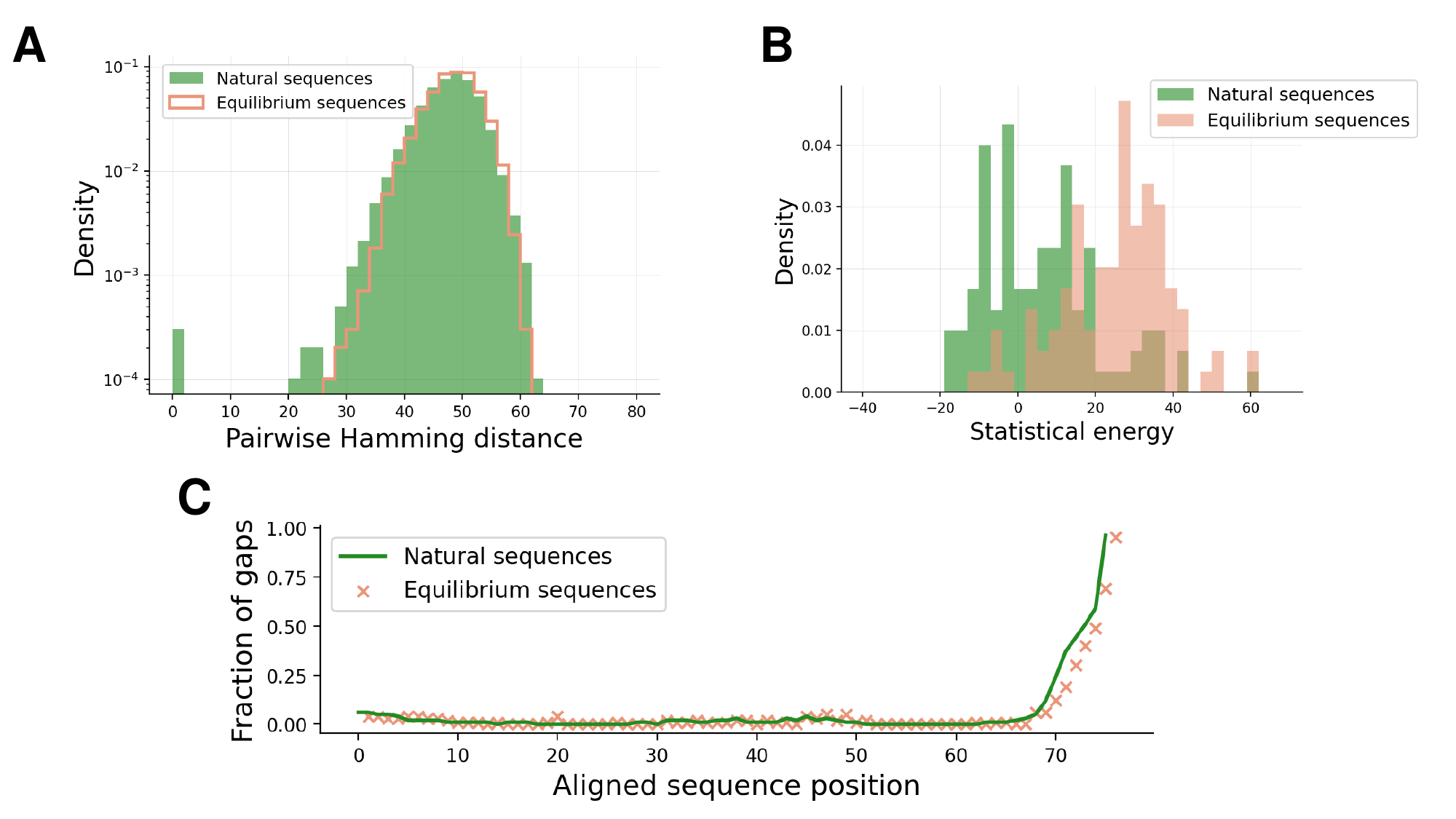}
\caption{{\bf Statistical properties of 100 equilibrium in silico sequences for the DBD family.}
(A)~Pairwise Hamming distance of natural
and equilibrium sequences.
(B) Energy of natural and equilibrium sequences.
(C) Fraction of gaps per position of natural and equilibrium sequences. The excess of
gaps at the extremities of the alignment is due to the alignment procedure.
}
\label{fig:s3}
\end{figure}

\begin{figure}[htbp]
\centering
\includegraphics[width=1.\linewidth]{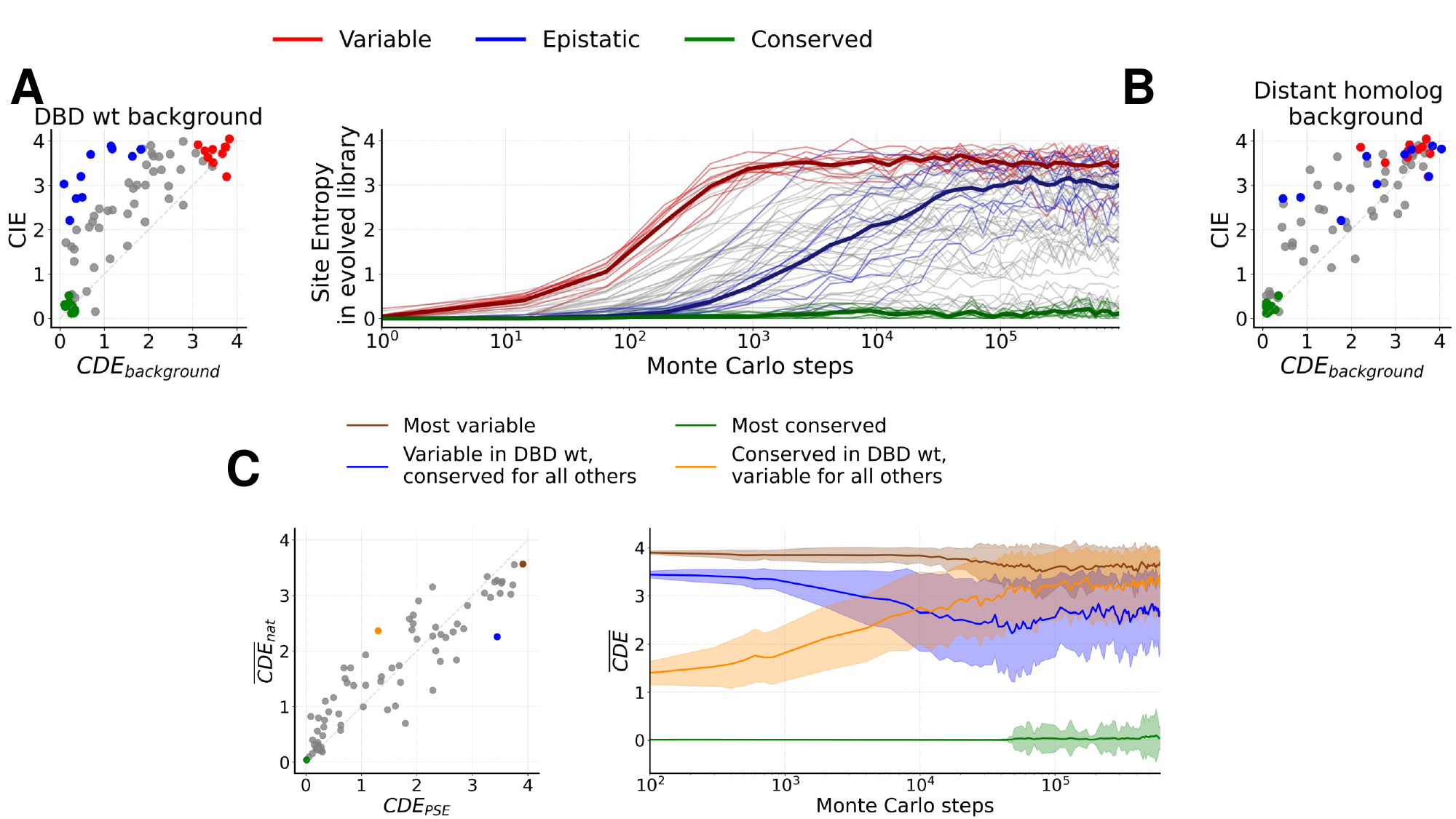}
\caption{{\bf In silico evolutionary dynamics of variable, conserved, and epistatic sites for the DBD family.}
(A) Left: classification of residues according to local variability (CDE) computed with the bmDCA model in the ancestral DBD sequence from Ref.~\cite{park2022epistatic} (wt) context and global variability (CIE) computed using the amino acid frequencies in natural proteins. We identify three categories of 10 sites each: conserved (green), mutable (red) and epistatic (blue). Right: Site entropy of a library constructed by 100 independent MCMC samplings initialized in the wt sequence, with the three different site categories highlighted; faded lines refer to each site whereas the three thick lines are the mean over the three categories. (B) Same scatter plot as in the left panel of A, but in the context of a distant homolog. The site colors are the same as for the DBD ancestral background, highlighting that site categories are reshuffled in a different context. (C) Left: Illustration of four prototypical sites in the alignment and scatter plot of their CDE computed in the wt context versus the mean CDE on all proteins of the DBD family. Right: CDE of the four sites identified in the left panel, averaged over the context generated by 100 independent MCMC trajectories (thick lines). The faded area indicates the standard deviation.}
\label{fig:s4}
\end{figure}

\begin{figure}[htbp]
\centering
\includegraphics[width=1.\linewidth]{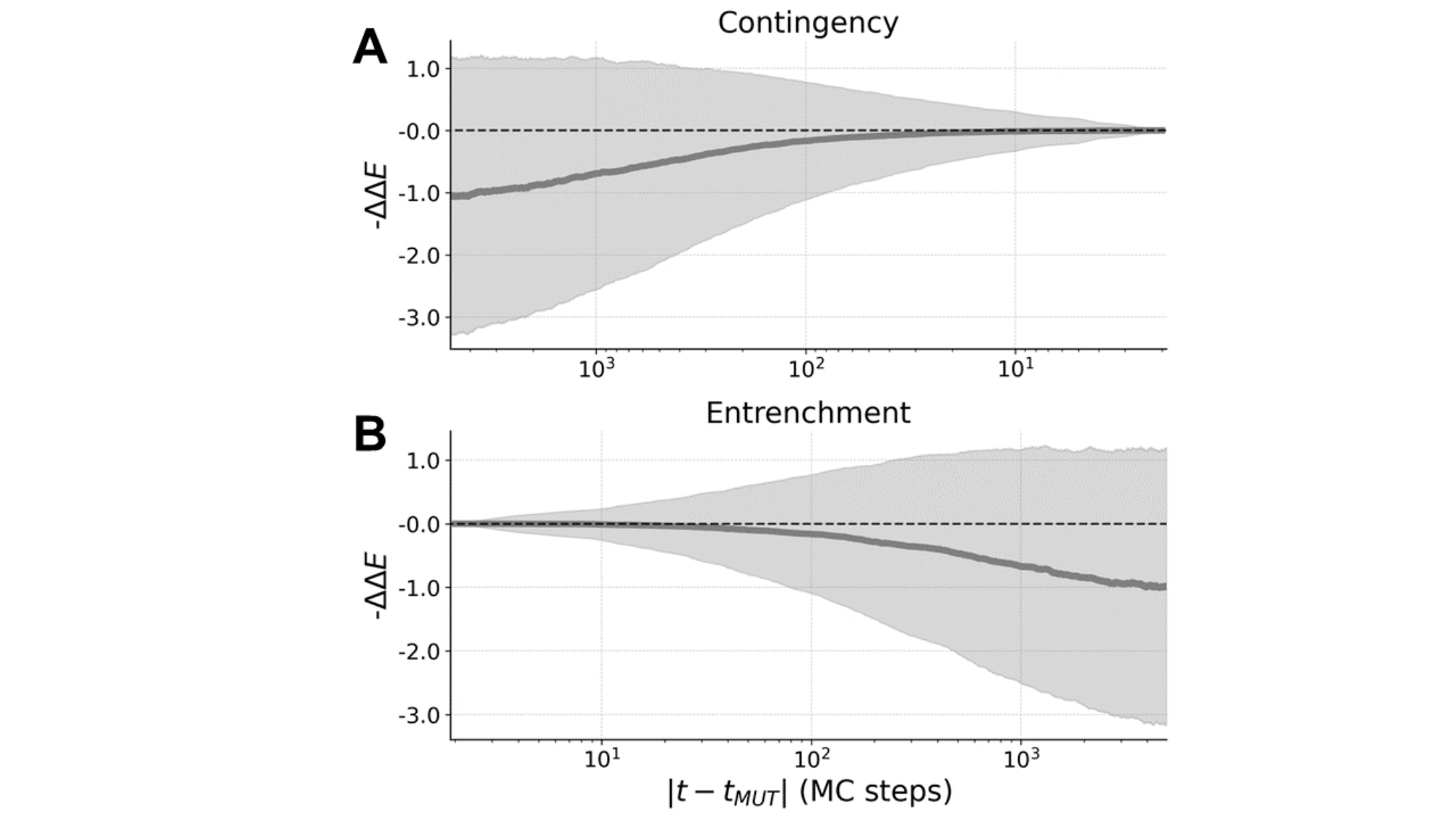}
\caption{{\bf Contingency and entrenchment from in silico evolution for the DBD family.}
We select 5000 mutations labeled by $aib$ (i.e. site $i$, aminoacid $a\to b$) along an evolutionary trajectory of the DBD family.
(A)~Contingency: average value of the fitness gain $-\Delta \Delta E$ between the original mutation $aib$ at time $\tmut$, and the mutational effect of introducing amino acid $b$ in site $i$ over a time window of 5000 MCMC steps before $\tmut$.
The faded area is the standard deviation. (B)~Entrenchment: 
average value of the fitness gain $-\Delta \Delta E$ between the original mutation $aib$ at time $\tmut$,
and the mutational effect of reverting to aminoacid $a$ over a time window of 5000 MCMC steps after $\tmut$. The faded area is the standard deviation.}
\label{fig:s5}
\end{figure}




\clearpage

\subsection{Dynamics of the local entropy in the absence of coevolutionary couplings}

\begin{figure}[htbp]
\centering
\includegraphics[width=1.\linewidth]{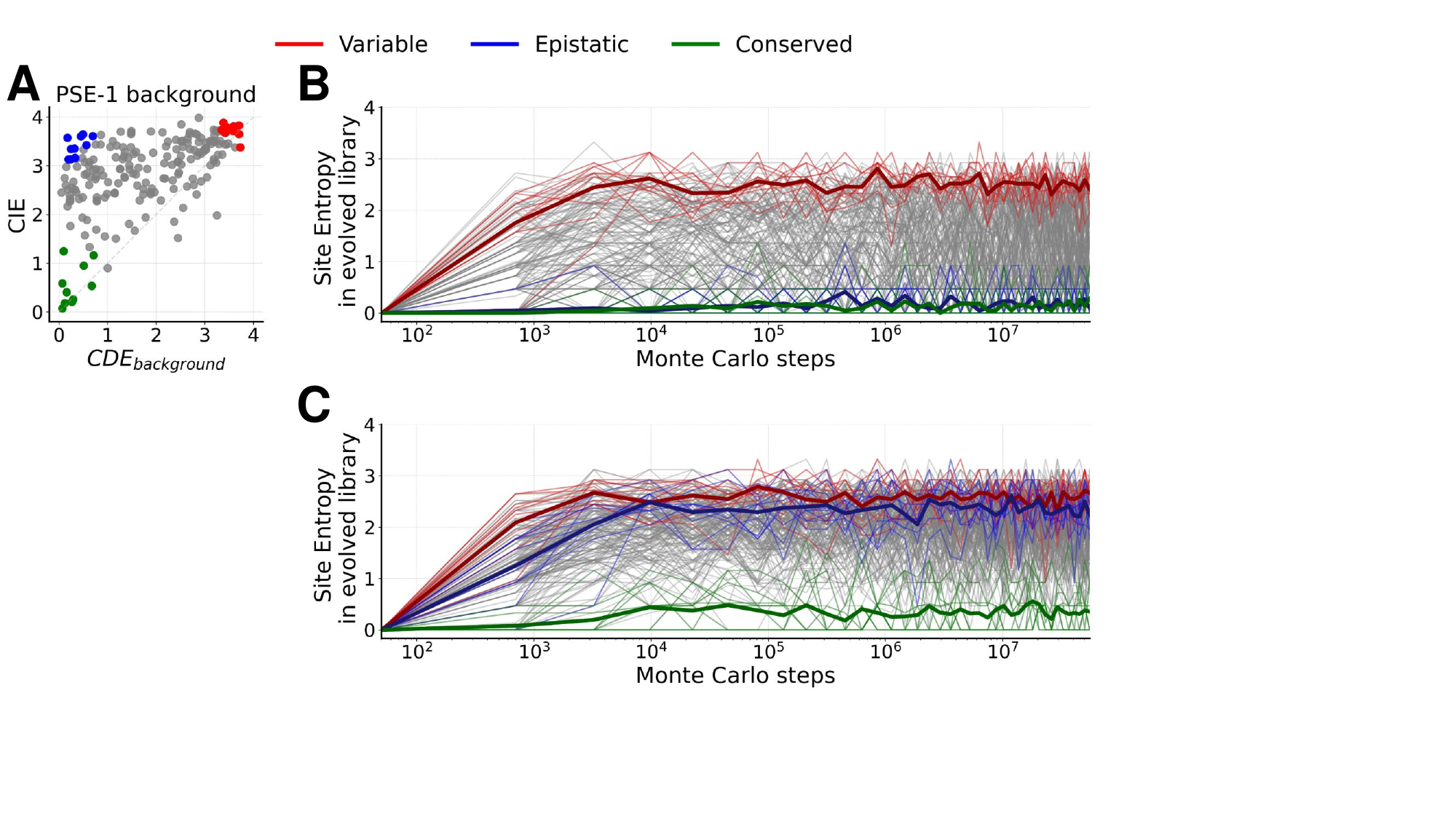}
\caption{{\bf In silico evolutionary dynamics of variable, conserved, and epistatic sites for the $\beta$-lactamase family using profile models.} To highlight the importance of the coevolutionary couplings, here we report the same analysis of Fig.~2 of the main text, but for a model without couplings, i.e. a profile model. The labels and color codes are identical to Figs.~2 and~\ref{fig:s4}. 
Panel B reports the results for a profile model that reproduces the context of PSE-1 as predicted by the bmDCA model, i.e.
we set $\pi_i(a_i) = P_{bmDCA}(a_i | \text{PSE-1})$ for each site and the global probability of the profile model is 
$P(a_1,\cdots,a_N) = \prod_{i=1}^N \pi_i(a_i)$.
Panel C reports the results for a profile model based on the natural amino acid frequencies, i.e. $\pi_i(a_i) = f_i(a_i)$.
As expected, we observe that in the absence of couplings, all site entropies evolve quickly to their equilibrium entropy in the profile model, which is
the CDE in the PSE-1 background in panel B, and the CIE in panel C.
 }
\label{fig:s6}
\end{figure}

\clearpage

\subsection{Classification of sites according to the DMS in the DBD family}

\begin{figure}[htbp]
\centering
\includegraphics[width=1.\linewidth]{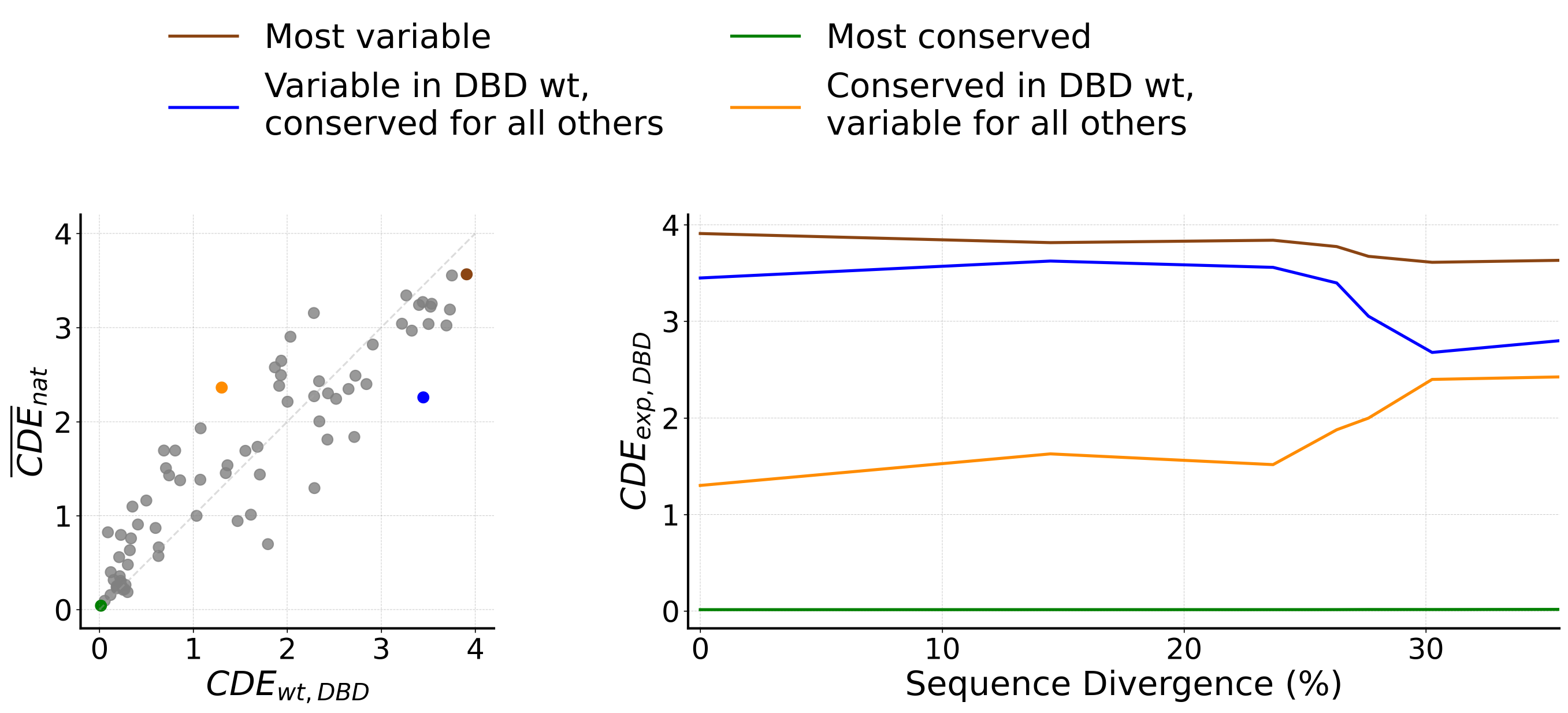}
\caption{{\bf Evolutionary dynamics of variable, conserved, and epistatic sites for the DBD experimentally studied ancestors.}
Left: Illustration of four prototypical sites, same figure as on the left of Fig.~\ref{fig:s4}C, repeated here for convenience.
Right: Instead of the in silico Monte Carlo chains used on the right of Fig.~\ref{fig:s4}C, we repeat the same calculation for the experimentally investigated DBD sequences, along the reconstructed phylogenetic lineage (seven sequences) 
starting from the ancestral protein and reaching {\it C. teleta} SR, i.e.~the upper purple branch in Ref.~\cite[Fig.1B]{park2022epistatic}.
We plot the 
 CDE of the four sites identified in the left panel, computed from the bmDCA model, as a function of the sequence divergence, starting from the ancestor (0\%) to the {\it C. teleta} SR (34\%). The evolution of the CDE is similar to the in silico trajectories
of Fig.~\ref{fig:s4}C.
  }
\label{fig:s7}
\end{figure}

\clearpage

\subsection{Single-mutation effects and codon bias}

\begin{figure}[htbp]
\centering
\includegraphics[width=0.55\linewidth]{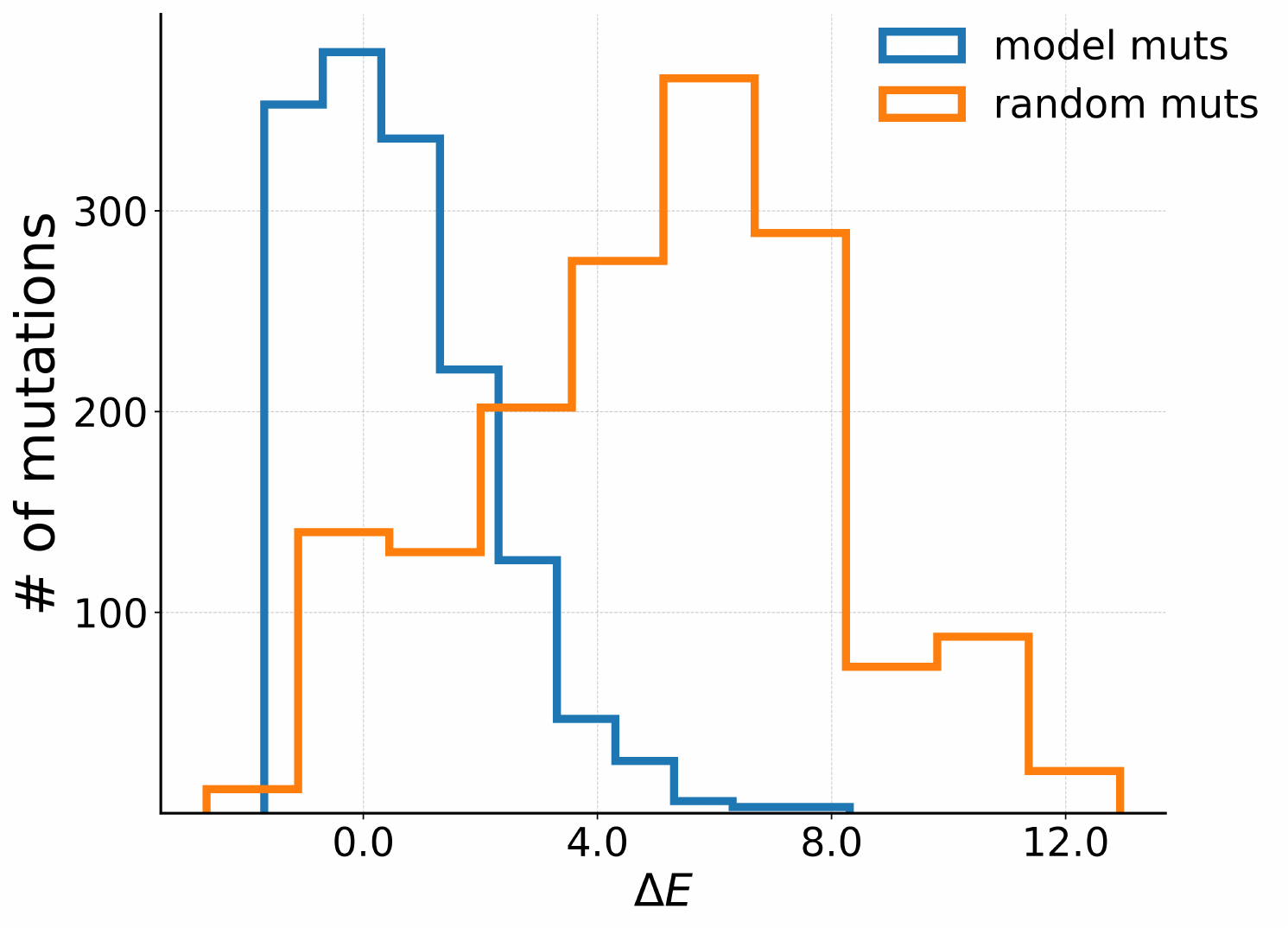}
\caption{{\bf Neutrality of mutational effects in simulated evolutionary dynamics.} 
Histogram of mutational effects of random point mutations (orange, not taking into account nucleotides) from DBD reconstructed ancestor vs. the histogram of in-silico observed mutations taking place over a sample of short trajectories around the same wt (blue, using the sampling procedure based on nucleotides). Higher $\Delta E$ indicates more deleterious mutations and vice-versa.
  }
\label{fig:s8}
\end{figure}

\begin{figure}[htbp]
\centering
\includegraphics[width=0.55\linewidth]{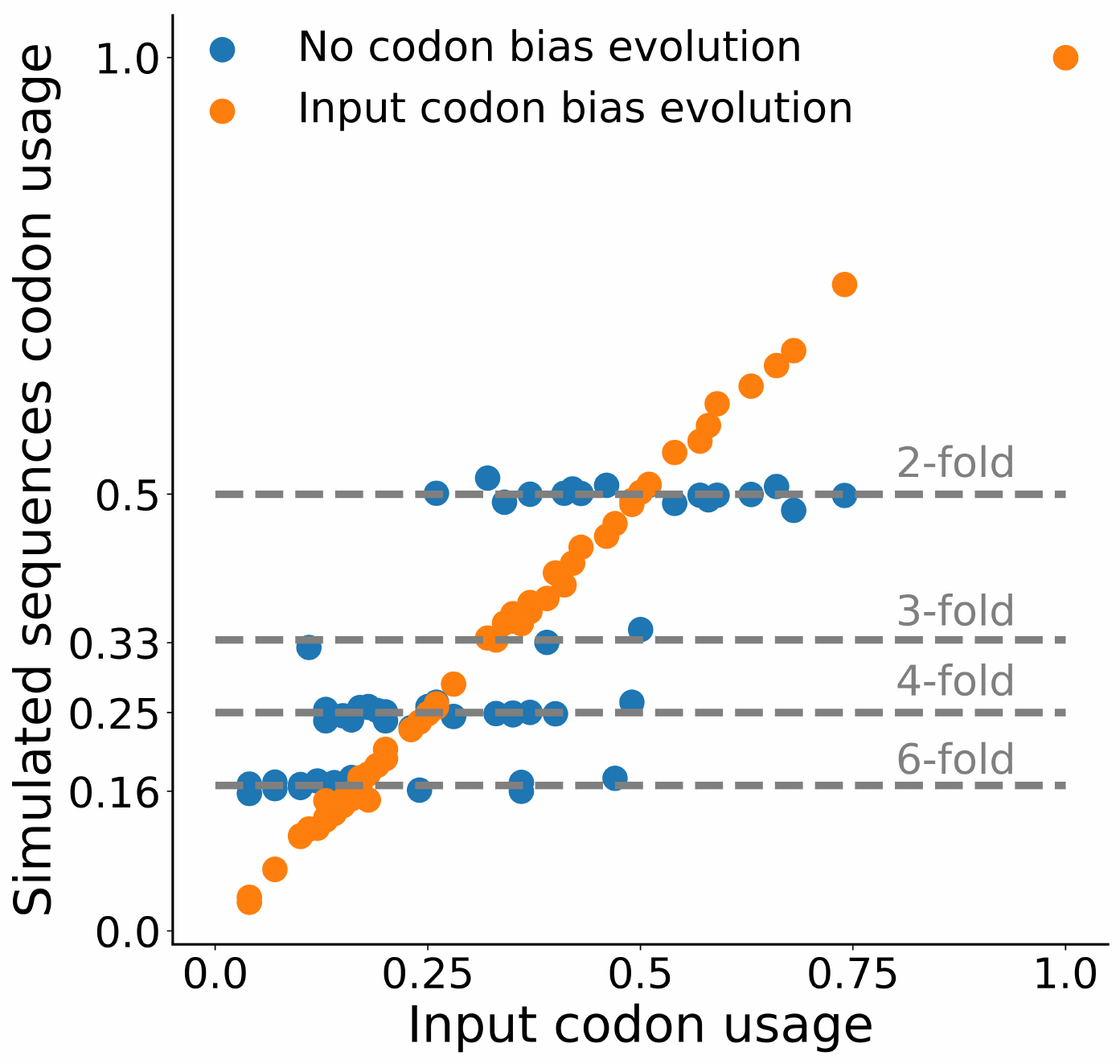}
\caption{{\bf Effect of the codon bias in sampling algorithm.} 
Scatter plot of E.coli codon usage vs. codon usage of in-silico sequences simulated with the E.coli codon bias (orange) or without codon bias (blue). All simulations were done for the DBD family. As we expect, sequences simulated with E.coli codon bias correctly reproduce E.coli codon usage. In contrast, if no codon bias is introduced, the codon usage resembles the n-fold degeneracies of genetic code: each codon $\bf{n}$ has a usage which is $1/[N(A({\bf{n}}))]$ where $N(A({\bf{n}}))$ indicates how many codons code the same amino acid of codon $\bf{n}$. 
  }
\label{fig:s9}
\end{figure}

\clearpage
\subsection{Role of insertions and deletions}

\begin{figure}[htbp]
\centering
\includegraphics[width=1.2\linewidth]{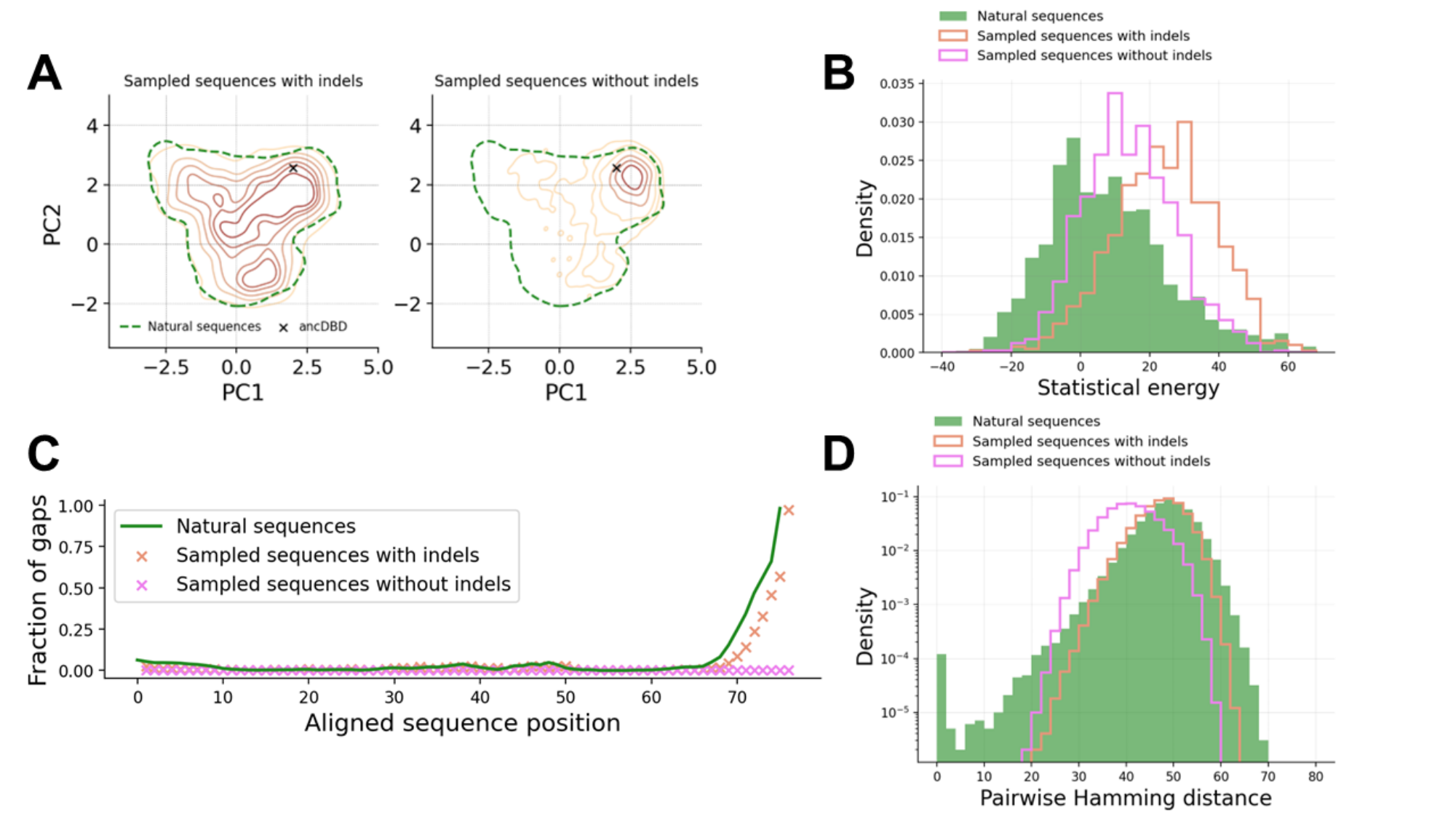}
\caption{{\bf Necessity of indels to retain generative properties of the sampling algorithm.} 
(A) Left: Projection over the two principal components (PCA) of natural sequences and sequences sampled with the standard algorithm (i.e. with indels). Right: Projection over the two principal components (PCA) of natural sequences and sequences sampled with a modified algorithm that excludes indels. In black the reconstructed ancestral sequence in which simulation has started. 
(B) Histogram of statistical energies for natural sequences (green), sequences sampled with indels (orange) and without indels (pink). 
(C) Fraction of gaps at every sequence position for natural sequences (green), sequences sampled with indels (orange) and without indels (pink). 
(D) Histogram of pairwise hamming distance among natural sequences (green), sequences sampled with indels (orange) and without indels (pink). All simulations were performed for the DBD family. 
  }
\label{fig:s10}
\end{figure}

\clearpage

\section{Contingency and entrenchment: a comparison with experimental data}

To further validate our evolutionary model, we test our results for contingency and entrenchment against experimental data. 
The authors of Ref.~\cite{park2022epistatic} reconstructed a phylogenetic lineage of a protein belonging to the DNA Binding Domain (DBD) family, from an ancient ancestor down to a present-day sequence, with a sequence divergence of about $35\%$ between the ancestor and the most recent sequence.
Then, they experimentally measured the effect of all mutations, i.e.~a DMS, around several protein sequences along the lineage, with variable sequence divergence. In line with expectations derived from our model, contingency and entrenchment were experimentally detected in those data. We can then quantitatively compare these experimental results with our in silico results.

\begin{figure*}[b]
\centering
\includegraphics[width=1.0\linewidth]{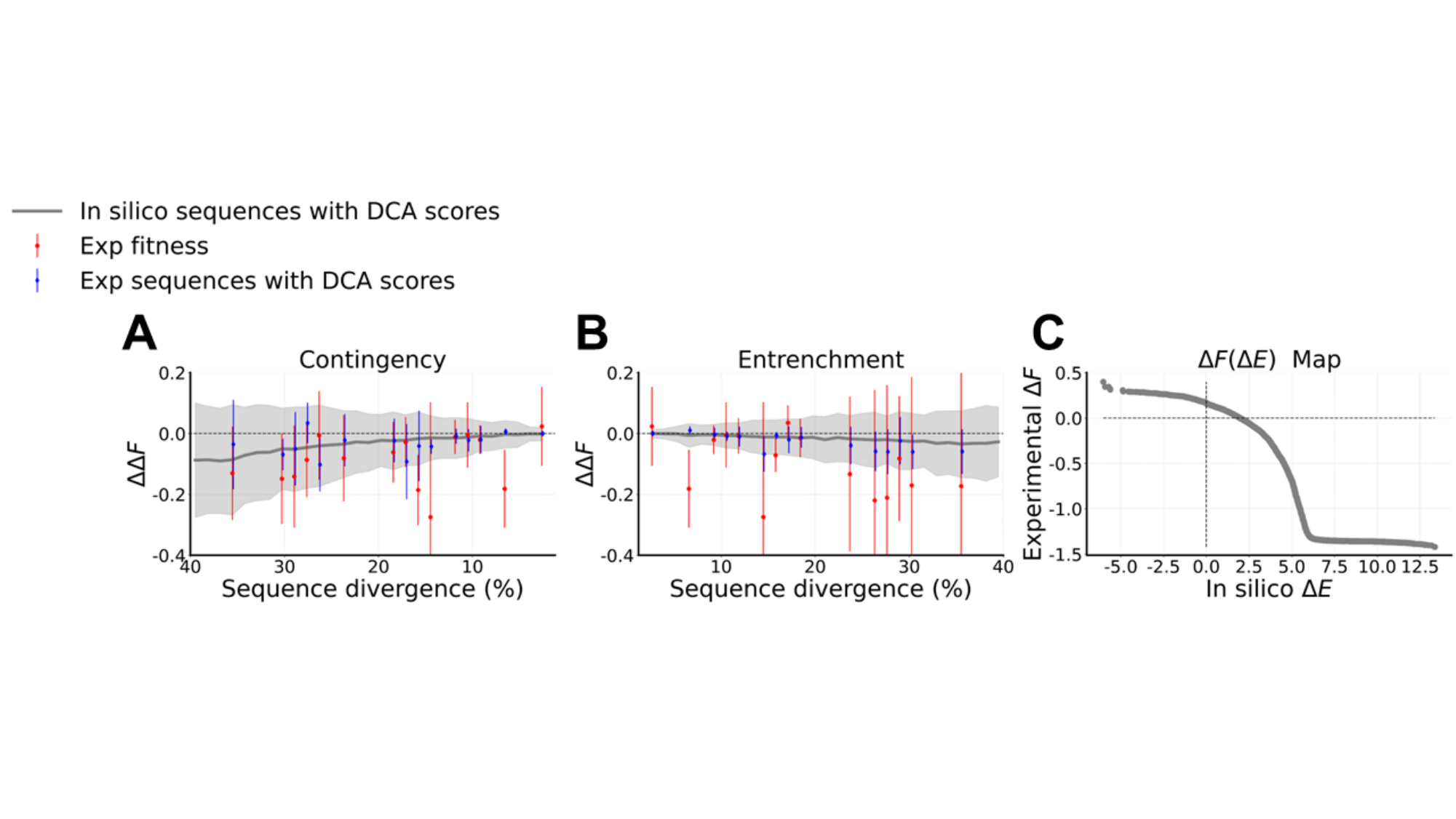}
\vskip-50pt
\caption{
{\bf Comparison between in silico and experimental results for the DBD family.}
(A)~Contingency: amino acids appearing for the first time over an evolutionary trajectory at step $\tmut$ are introduced back in all sequences at $t<\tmut$ and the difference of mutational effect at $t$ and $\tmut$ is displayed as a function of sequence divergence between the two reference sequences. (B)~Entrenchment: amino acids appearing for the last time over an evolutionary trajectory at step $\tmut$ are introduced in all sequences at $t>\tmut$ and the difference of mutational effect at $t$ and $\tmut$ is displayed as a function of sequence divergence between the two reference sequences. Results in (A) and (B) compare in silico sequences along an ensemble of trajectories generated by our evolutionary model (grey), experimental sequences with a score based on the experimental fitness (red) and experimental sequences with DCA scores based on the in silico model energy (blue). Faded area and error bars are the standard deviation of all mutations happening at that sequence divergence. (C) Visualization of the mapping $\Delta F =\phi(\Delta E)$ that is used to convert DCA energies into experimental fitness, constructed as described in the text.
}
\label{fig:4}
\end{figure*}

First of all, we note that the DCA mutational score is derived from naturally occurring proteins, and it is therefore a generic proxy for the general fitness of a protein along multiple conditions and phenotypes that are relevant to natural selection.
Hence, to compare with a specific experiment,
we need to map the experimental fitness score $\Delta F$ considered in that experiment to our scores $\Delta E$. To do so, following previous work~\cite{figliuzzi2016coevolutionary,barrat2016improving}, we consider the ensemble of single mutational effects $\Delta F$ measured in experiments (i.e., we collect all DMSs from Ref.~\cite{park2022epistatic}) and for each mutation (in its context) we compute the corresponding $\Delta E$ from the model. We then independently sort both lists of $\Delta E$ and $\Delta F$, and plot the sorted lists against each other in Fig.~\ref{fig:4}C (considering that high fitness corresponds to low statistical energy). In this way, we obtain a monotonous function that estimates the non-linear mapping $\Delta F = \phi(\Delta E)$ and converts energy variations into fitness variations for this particular experiment.

\begin{figure*}[t]
\centering
\includegraphics[width=1.0\linewidth]{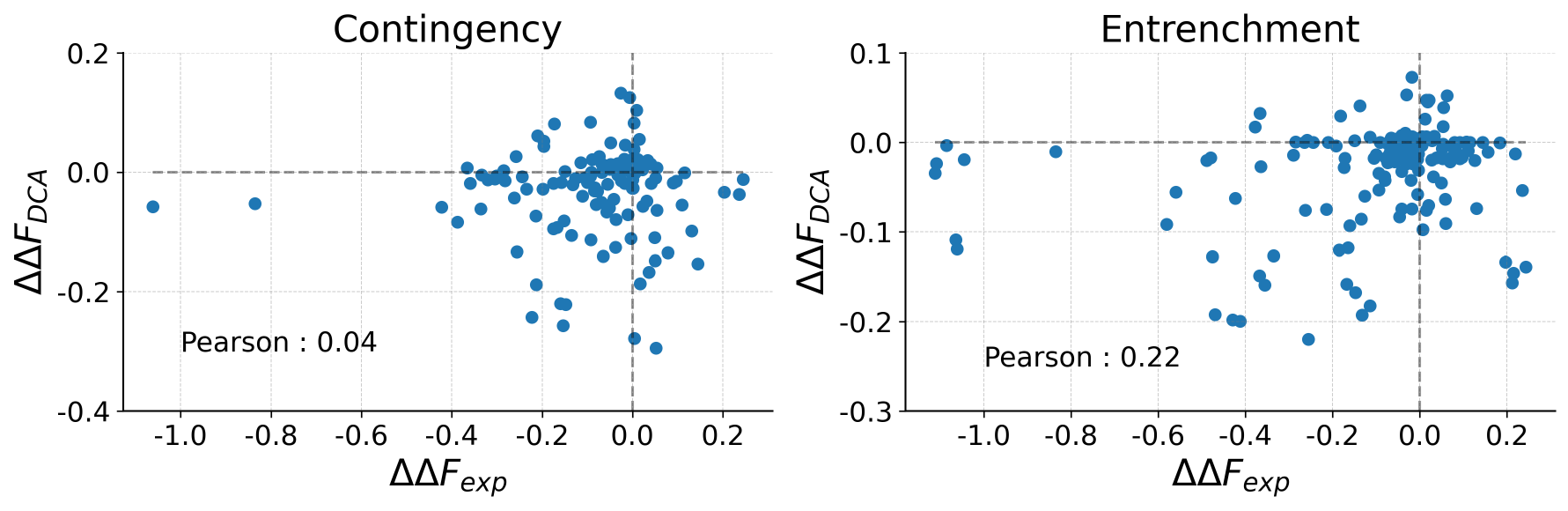}
\caption{
{\bf Comparison between in silico and experimental change in mutational effects over time for the DBD family.}
We take all the single $\Delta\Delta F_{exp}$ irrespectively of the Hamming distance between the experimental sequences and we compare them with the same changes of mutational effects $\Delta\Delta F_{DCA}$ predicted by the model, showing the Pearson correlation coefficient. We do the analysis separately for changes in mutational effects linked to contingency and entrenchment. The $\Delta\Delta F_{DCA}$ is obtained by the mapping $\Delta F =\phi(\Delta E)$ that is used to convert DCA energies into experimental fitnesses, constructed as described in the text.
}
\label{fig:s12}
\end{figure*}

Given $\phi(\Delta E)$, we repeat the analysis of contingency and entrenchment discussed in Fig.~\ref{fig:s3} as applied to the DBD family, converting the $\Delta E$ into $\Delta F$ using the non-linear mapping. Also, to allow for a direct comparison with experiments, we plot simulation results as a function of sequence divergence between the sequences at times $\tmut$ and $t+\tmut$.
The in silico results are reported in Fig.~\ref{fig:4}A for contingency and in Fig.~\ref{fig:4}B for entrenchment (grey lines and shades). Note that these results are obtained on typical evolutionary trajectories generated by the model, that are uncorrelated from the specific lineage studied experimentally in~\cite{park2022epistatic}.

Next, we superimpose our in silico results with the experimental measurements of $\Delta\Delta F$ obtained by comparing the effect of the same mutation, taken from any pair of DMS around two distinct wildtypes along the phylogenetic lineage in Ref.~\cite{park2022epistatic}, as a function of the divergence between these wildtypes.
The red points and standard deviations in Figs.~\ref{fig:4}A-B show, with exception of few outliers, a good quantitative agreement with our in silico results.
To complete the picture, we also compute the $\Delta E$ from the model around the same wild types used in the experiment, convert them into $\Delta F$ using the non-linear mapping $\phi$, and perform the same analysis (blue points), obtaining consistent results.
We thus see that our evolutionary model captures the trend in experimental data, thus confirming its usefulness as a tool to recapitulate and expand experimental evolution results.  

For the sake of completeness, we also compare the single  $\Delta \Delta F$ of mutations for experimental and in silico fitness, irrespectively of the distance between the two divergent reconstructed homologs (Fig.~\ref{fig:s12}). We see that our model does not quantitatively predict a $\Delta \Delta F$ given the two sequences and the mutation taking place among them. There are two main motivations for this lack of predictive power. Firstly, as we look more carefully at the scale of  $\Delta \Delta F$ (Fig.~\ref{fig:s12}) as compared to that of  $\Delta F$ (Fig.~\ref{fig:4}C) we see that contingency and entrenchment are very small effects (in a context of much higher experimental variability) that can be appreciated only at a sequence divergence of $30-40\%$ and are consequently pretty hard to model. Secondly, our statistical model is by construction reflecting the global landscape of highly divergent ($70 - 80\%$ of Hamming distance) natural sequences belonging to the same family, hence it cannot have the appropriate resolution for measuring a fine scale change in $\Delta F$ over time caused by a single mutation.

\clearpage
\bibliographystyle{mioaps}
\bibliography{pnas-sample}

\end{document}